\documentclass[conference]{IEEEtran}
\IEEEoverridecommandlockouts
\usepackage{cite}
\usepackage{amsmath,amssymb,amsfonts}
\usepackage{algorithmic}
\usepackage{graphicx}
\usepackage{textcomp}
\usepackage{xcolor}
\usepackage{subcaption}
\usepackage{cleveref}
\usepackage{url}
\usepackage{todonotes}

\def\BibTeX{{\rm B\kern-.05em{\sc i\kern-.025em b}\kern-.08em
    T\kern-.1667em\lower.7ex\hbox{E}\kern-.125emX}}
\begin{document}

\title{Analysis of parallel I/O use on the UK national supercomputing service, ARCHER using Cray's LASSi and EPCC SAFE
\thanks{This work was supported by the UK National Supercomputing Service, ARCHER (http://www.archer.ac.uk); funded by EPSRC and NERC.}
}

\author{
\IEEEauthorblockN{Andrew Turner}
\IEEEauthorblockA{\textit{EPCC} \\
\textit{The University of Edinburgh}\\
Edinburgh, UK \\
a.turner@epcc.ed.ac.uk}
\and
\IEEEauthorblockN{Dominic Sloan-Murphy}
\IEEEauthorblockA{\textit{EPCC} \\
\textit{The University of Edinburgh}\\
Edinburgh, UK \\
d.sloan-murphy@epcc.ed.ac.uk}
\and
\IEEEauthorblockN{Karthee Sivalingam} 
\IEEEauthorblockA{\textit{Cray European Research Lab} \\
Bristol, UK \\
ksivalinga@cray.com}
\and
\IEEEauthorblockN{Harvey Richardson}
\IEEEauthorblockA{\textit{Cray European Research Lab} \\
Bristol, UK \\
harveyr@cray.com}
\and
\IEEEauthorblockN{Julian Kunkel}
\IEEEauthorblockA{\textit{Department of Computer Science} \\
\textit{University of Reading}\\
Reading, UK \\
 j.m.kunkel@reading.ac.uk}
}

\maketitle

\begin{abstract}
In this paper, we describe how we have used a combination of the LASSi tool (developed by Cray) and the SAFE software (developed by EPCC) to collect and analyse Lustre I/O performance data for all jobs running on the UK national supercomputing service, ARCHER; and to provide reports on I/O usage for users in our standard reporting framework. We also present results from analysis of parallel I/O use on ARCHER and analysis on the  potential impact of different applications on file system performance using metrics we have derived from the LASSi data. We show that the performance data from LASSi reveals how the same application can stress different components of the file system depending on how it is run, and how the LASSi risk metrics allow us to identify use cases that could potentially cause issues for global I/O performance and work with users to improve their I/O use. We use the IO-500 benchmark to help us understand
how LASSi risk metrics correspond to observed performance on the ARCHER
file systems. We also use LASSi data imported into SAFE to identify 
I/O use patterns associated with different research areas, understand
how the research workflow gives rise to the observed patterns and 
project how this will affect I/O requirements in the future. Finally,
we provide an overview of likely future directions for the continuation
of this work.
\end{abstract}

\begin{IEEEkeywords}
Supercomputers, High performance computing, Parallel architectures, Data storage systems, Performance analysis 
\end{IEEEkeywords}

\section{Introduction}

I/O technologies in supercomputer systems are becoming increasingly complex and diverse.
For example, a recent trend has been to add a new kind of high-performance but limited capacity I/O to supercomputing systems---often
referred to as \textit{burst-buffer} technologies. Recent examples include Intel Optane~\cite{halfacree2015intel} and Cray DataWarp~\cite{henseler2016architecture}. These technologies typically provide orders of magnitude more performance, both in terms of I/O
bandwidth and I/O operations per second, at the expense of total storage capacities.

To help establish the potential impact of such novel technologies within the HPC sphere, we need to revisit and update our data on the typical I/O requirements of modern
applications. 

There are many factors which affect the I/O behavioural requirements of any scientific application, 
and these factors have been changing rapidly in recent years. For example, the
ratio of network performance to node-level performance tends to influence how much work each
node needs to perform. As the node-level performance tends to grow faster than the network-level
performance, the trend is for each node to be given more work, often implying larger I/O requirements per node.
The complexity of the interactions between performance behaviour and system development are discussed by Lockwood et al~\cite{lockwood2018year}.
They investigate the performance behaviour from the perspective of applications and the file system quantifying the performance development over the course of a year.
Due to these changes, we cannot
rely on conventional wisdom, nor even older results, when understanding current I/O requirements on HPC systems. Instead, we need up-to-date, good quality data with which to reason and inform our assumptions of current systems and predictions of future systems.

In this study, we have used 
ARCHER\footnote{\url{http://www.archer.ac.uk}}---the UK's national supercomputer---as an example of a high-end supercomputer. ARCHER reached \#19 in the Top500 upon its launch in
2013. It is a 4,920 node Cray XC30, and consists of over 
118,000 Intel Ivy Bridge cores, with two 2.7 GHz, 12-core E5-2697 v2 CPUs per node. 4,544 of the 4,920 nodes have 64 GiB per node (2.66 GiB per core),
while the remaining 376 `high memory' nodes have 128 GiB each (5.32 GiB per core). The ARCHER production service has three Lustre file systems
each based on a Cray Sonexion 1600 appliance. Two file systems have
12 OSS and one file system has 14 OSS. Each OSS is a Seagate Sonexion
1600 OSS controller module, 1 x Intel Xeon CPU E5-2648L @ 1.80GHz, 32GB
memory. Each OSS has 40 discs, 4 OSTs per OSS, 10 discs per OST. These
10 discs are in RAID6, i.e. 8+2. There are also a number of hot spares and
RAID and ext3 journaling SSDs on each OSS. Each disc is a 4TB
SEAGATE ST4000NM0023 (Constellation ES.3 - 3.5" - SAS 6Gb/s - 7,200 rpm).
There is one MDS and one backup MDS per file system. Each MDS is a
Cary Sonexion 1600 MDS controller module, 2 x Intel(R) Xeon(R) CPU
E5-2680 @ 2.70GHz Each of the 3 MDTs comprise 14 discs in RAID10.
Each disc is a 600GB SEAGATE ST9600205SS (Enterprise Performance 10K
600 GB - 2.5" - SAS 6Gb/s - 10,000 rpm). Each client accesses the
three file systems via 18 LNet router nodes internal to the ARCHER
system. Each of the three file systems are attached to 10, 10 or 14 router nodes respectively; some
router nodes service more than one path. This is complex, involving
overlapping primary and secondary paths, however, the rule that affects
performance is that the primary LNet path is configured so that all
clients access 3 OSS nodes via 2 LNet router nodes. MDS nodes are accessed from the clients via 2 LNet router nodes each.

HPC applications scheduled to run on ARCHER have to share resources,
in particular the file system and network. Even though these shared
resources are built to scale well and provide high performance, they
can become a bottleneck when multiple applications stress them at the same
time. Occasionally the applications also use these shared resources
inefficiently, which may impact other applications using the same
resource.

Users expect applications to perform consistently in a time frame, i.e., the overall runtime for a given job does not vary excessively. 
Often time limits are chosen such that slowdown can cause jobs to fail. 
However, from time to time users would report that their applications were running slower than expected or interactive file system response was sub-optimal. Based on this feedback, we set out to analyse all of the applications running on ARCHER for their current
I/O usage, to try to understand the variability of I/O performance on the system and its link to running applications. 
In contrast to other studies (that typically profile the I/O use of a small number of benchmark applications), we are 
sampling the I/O usage of \textit{every} job run on ARCHER in the analysis 
period. Thus our data should complement those from previous studies.

Most monitoring tools  \cite{Berkeley2009DeployingSF},
\cite{Shipman2010LessonsLI}, 
\cite{uselton2010file}, \cite{LMT}, \cite{roland_laifer} and
\cite{miller_article} only provide raw I/O statistics 
of file systems or applications. 
UMAMI~\cite{Lockwood:2017:URG:3149393.3149395}  and MELT~\cite{DBLP:journals/corr/BrimL15} add features for slowdown analysis but require expertise. 
Previous work introduced metrics such as I/O severity~\cite{Uselton2013AFS} and File System Utilisation(FSU)~\cite{Mendez:2017:API:3101112.3101260} for studying I/O and application slowdown.
We have developed a non-invasive framework where it is easy to identify applications with unusual I/O behaviour, and by targeting application interactions with the file system. The following sections describe this framework along with insights gained from running IO-500 benchmarks and detail the I/O patterns observed by the data analysis.

\section{Tools and Methodology}

This section first introduces the tools used to monitor the I/O utilisation and to relate them with user jobs.
To validate the behavior of this approach on a well known pattern, we utilise the IO-500 benchmark.

\subsection{LASSi}

LASSi (Log Analytics for Shared System resource with instrumentation)~\cite{lassi} was developed by the Cray Centre of Excellence (CoE) for ARCHER to provide system staff with the ability to find and understand contention in the file system.

LASSi is a tool to analyse the slowdown of applications due to the shared Lustre file system usage.
It provides HPC system support staff the ability to monitor and profile the I/O usage of applications over time. LASSi uses a metric-based approach to study the 
quantity and I/O quality. Metrics describe the risk of slowdown of applications at any time and also identifies the applications that cause such high risks. This information is then made available to the user or application
developer as appropriate.

LASSi was originally planned to be an extension of work undertaken by Diana Moise of Cray on the HLRS system \cite{Moise}. 
This work defined aggressor and victim jobs 
\emph{running at the same time}. 
Grouping applications based on 
the exact command line used, the study defined slowdown as a deviation from the mean
run times by 1.5 times or more. This study did not use any I/O or network statistics but was attempting to spot correlations in job runtimes.

\emph{Victim} detection was based on observing applications that run slower than the average
run time for an application group. \emph{Aggressor} detection was based on applications that overlap with the \emph{victims}. 
The \emph{Victim} and \emph{Aggressor} model based on concurrent running fails to provide useful insights when we move to a system like ARCHER, which is at a scale where there are always a large number of applications running.

In ARCHER, user reports of slowdown are usually addressed by analysing
the raw Lustre statistics, stored in a MySQL database called LAPCAT
(developed by Martin Lafferty from the onsite Cray systems team).
LAPCAT provides the following Lustre I/O statistics from each compute
node over time:

\begin{itemize}
    \item \textbf{OSS:} \emph{read\_kb, read\_ops, write\_kb,
    write\_ops, other}
    \item \textbf{MDS:} \emph{open, close, mknod, link, unlink,
    mkdir, rmdir, ren, getattr, setattr, getxattr, setxattr,
    statfs, sync, sdr, cdr}
\end{itemize}

Before LASSi, mapping the Lustre statistics to application runs and
looking for patterns using LAPCAT was a prohibitively long time to
investigate.

We designed LASSi to use defined metrics that indicate problematic
\emph{behaviour} on the Lustre file systems. Ultimately, we have
shown that there is less distinction between Victims and Aggressors.
An alternative explanation, supported by the LASSi-derived data, is
that so-called Victims are simply using the Lustre file system
more heavily than so-called Aggressors. 

Application run time depends on multiple factors such as 
compute  clock  speed,  memory  bandwidth,  I/O  bandwidth, network bandwidth and scientific configuration (dataset
size  or  complexity). LASSi aims only to model application run time 
variation  due to I/O.

\subsection{Risk-Metric Based Approach}

These metrics are motivated by the fact that we expect users will report slowdown only when their application run takes longer than usual. 
We focus on I/O as the most likely cause of unexpected application slowdown and
begin with the assumption that, in isolation, slowdown only happens 
when an application does more I/O than expected (for example, due to configuration or code change) or when an application has an unusually high resource requirement than normal at a time when the file system is busier than usual. 

To characterise situations that cause slowdown means considering raw I/O rate, metadata operations and quality (size) of I/O operations. For example, Lustre file system usage is optimal when at least 1 MB is read or written 
for each operation (\emph{read\_ops} or 
\emph{write\_ops}). 

The central metadata server can sustain a certain rate of metadata operations, above which any metadata request from any application or group of applications will cause slowdown. 
To provide the type of analysis required, LASSi must comprehend this complex job mix of different applications with widely different read/write patterns, the metadata operations running at the same time and how these interact and affect each other. This requirement informs the definition of the LASSi metrics.

\subsection{Definition of Metrics}

Firstly, we define metrics that indicate quantity and I/O quality operations by an application run. 
We first define the risk for any \emph{OSS} or \emph{MDS} operation $x$
on a file system \textit{fs} as
\begin{equation*}
risk_{fs}(x) = \frac{x-\alpha \cdot \text{avg}_{fs}(x)}{\alpha \cdot \text{avg}_{fs}(x)} 
\end{equation*}
where the averages are over the raw file system statistics and $\alpha$ is a scaling factor, set to 2 for this analysis.
The risk metric measures
the deviation of Lustre operations from the (scaled) average on a file system.
A higher value indicates higher risk of slowdown to a file system.
To simplify the representation for the user, the risk for metadata and data operations aggregate various types of operations into one value:

\begin{equation*}
\begin{split}
risk_{oss} &= risk_{read\_kb} + risk_{read\_ops} +  \\
&risk_{write\_kb} + risk_{write\_ops} + risk_{other}
\end{split}
\end{equation*}

\begin{equation*}
\begin{split} 
risk_{mds} &= risk_{open} + risk_{close} + risk_{getattr} + risk_{setattr} + \\
&risk_{mkdir} +  risk_{rmdir} + risk_{mknod} + risk_{link} + \\
&risk_{unlink} + risk_{ren} +  risk_{getxattr} + risk_{setxattr} + \\
&risk_{statfs} + risk_{sync} + risk_{cdr} + risk_{sdr}
\end{split}
\end{equation*}
\emph{Risks} for individual operations are added only if the value is greater than zero;
as any negative risks are ignored since this would correspond to the situation where the I/O was less than the average. The total risk on
a file system at a given time is the sum of all application risks.

For some metadata operations, the averages are closer to zero and this can cause the \emph{risk} metrics
to become very large. We still want to measure and identify applications that do exceptional metadata 
operations like creating thousands of directories per second. For these metadata operations, 
we use $\beta$-scaled average of the sum of all metadata operations to measure risk, where
$\beta$ is usually set to 0.25. Both $\alpha$ and $\beta$ are used to set the lower limit for defining the risks and this can be configured based on experience.

The above metric measures the quantity of I/O operations, but not the quality. 
On Lustre, 1\,MB aligned accesses are the optimal size per operation. 
To define a measure of the quality reads and writes, we define the following metrics:
\begin{equation}
read\_kb\_ops = \frac{read\_ops \cdot 1024}{read\_kb} 
\end{equation}
\begin{equation}
write\_kb\_ops = \frac{write\_ops \cdot 1024}{write\_kb}
\end{equation}

The read or write quality is optimal when (respectively) $read\_kb\_ops =1$ or $write\_kb\_ops=1$.
A value of $read\_kb\_ops >>1$ or $write\_kb\_ops>>1$ denotes poor quality read and writes.
The total ops metric on
a file system at a given time is sum of all application ops metric with $risk_{oss} > 0$ (ignoring applications with low quantity of I/O).
In general, \emph{risk} metrics measures the quantity of I/O and 
\emph{ops} metrics measures the quality.

\begin{figure}[tbp]
\centerline{\includegraphics[scale=0.38]{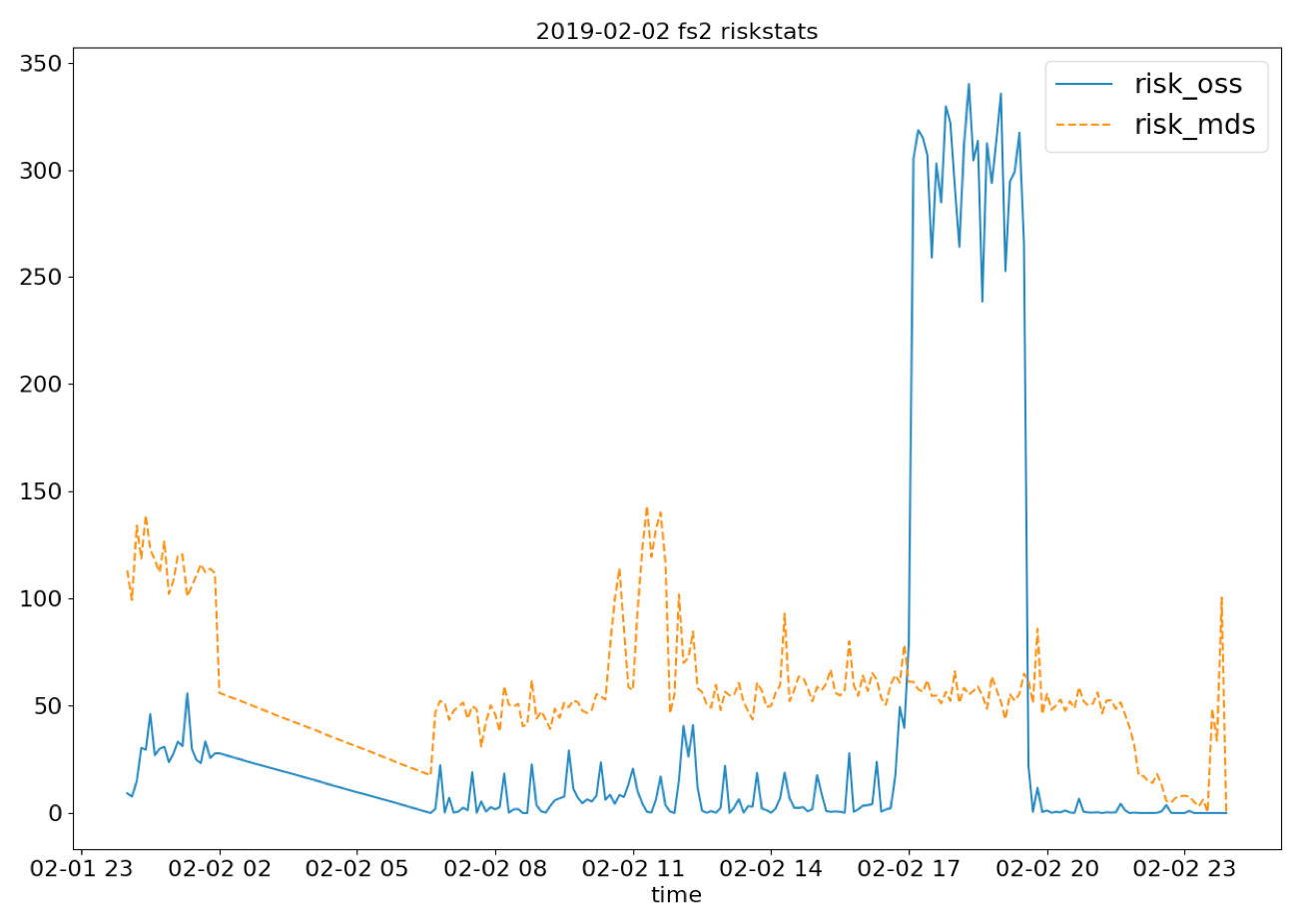}}
\caption{Sample report showing the risk to file system fs2 over
24 hours.}
\label{fig-lassi-risk}
\end{figure}

\begin{figure}[tbp]
\centerline{\includegraphics[scale=0.38]{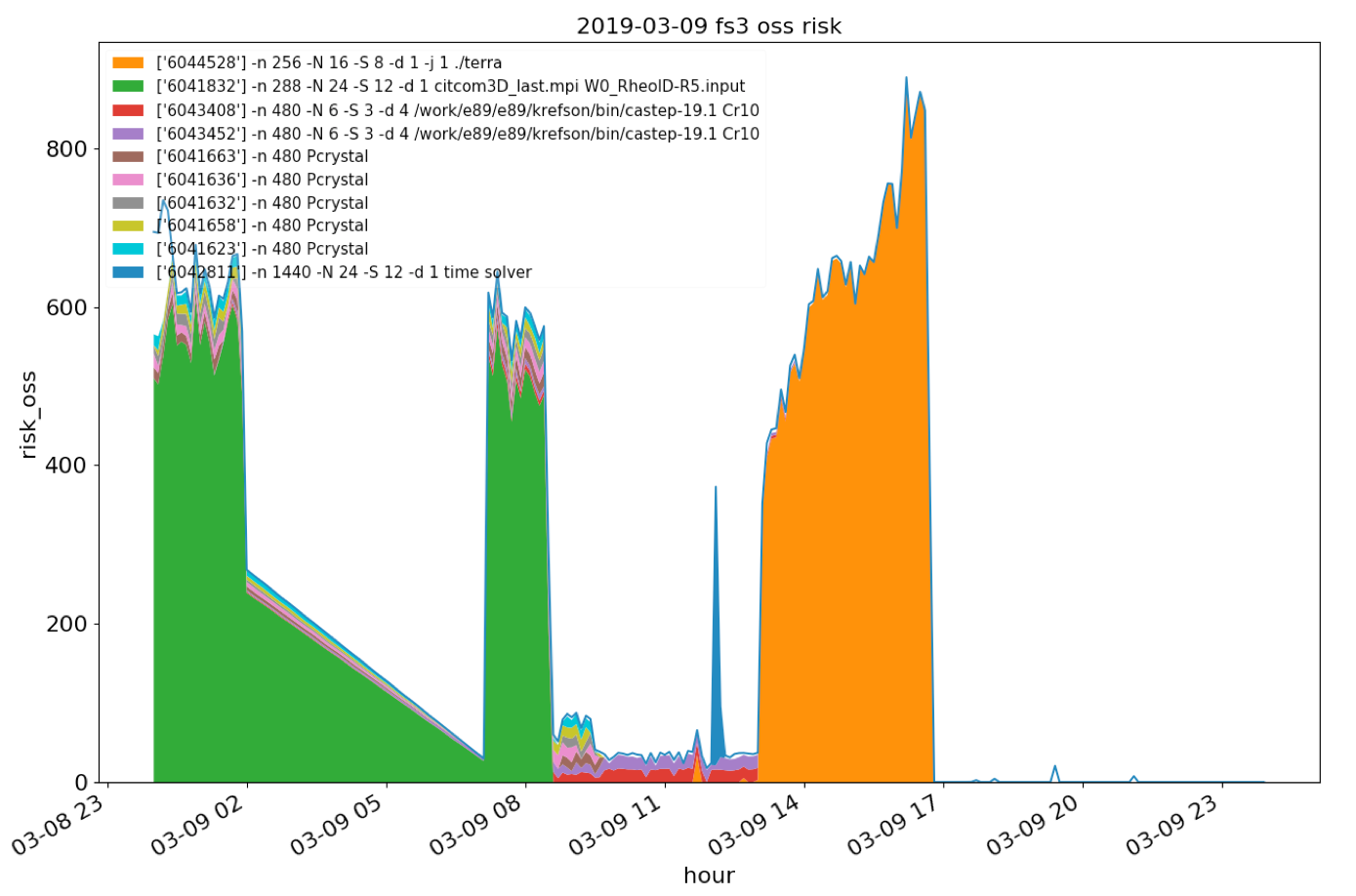}}
\caption{Sample report showing the OSS risk to file system fs2 over
24 hours with applications that are contributing to
the risk.}
\label{fig-lassi-oss-risk}
\end{figure}

\begin{figure}[tbp]
\centerline{\includegraphics[scale=0.38]{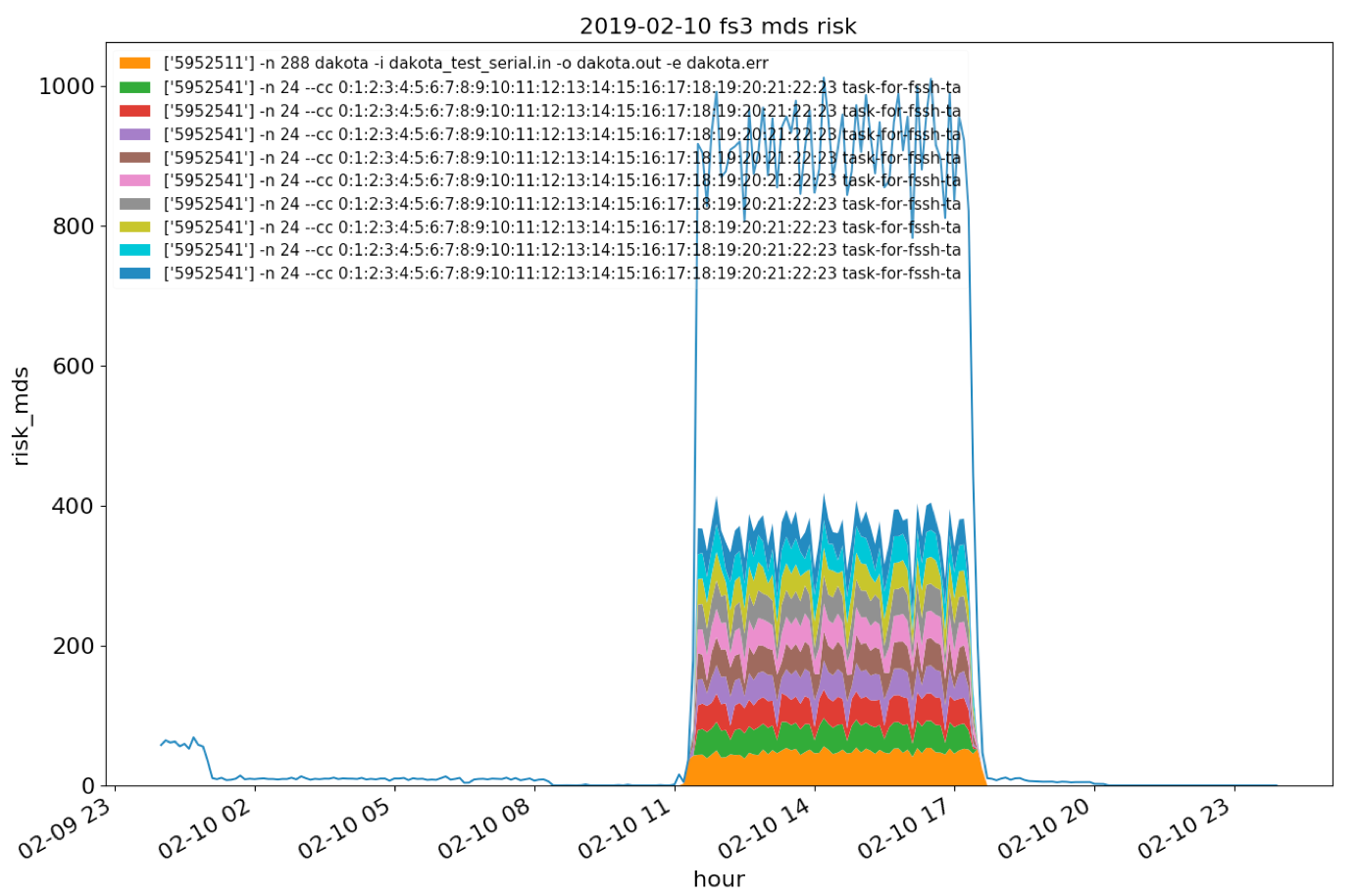}}
\caption{Sample report showing the MDS risk to file system fs2 over
24 hours with applications contributing to
the risk.}
\label{fig-lassi-mds-risk}
\end{figure}

\begin{figure}[tbp]
\centerline{\includegraphics[scale=0.38]{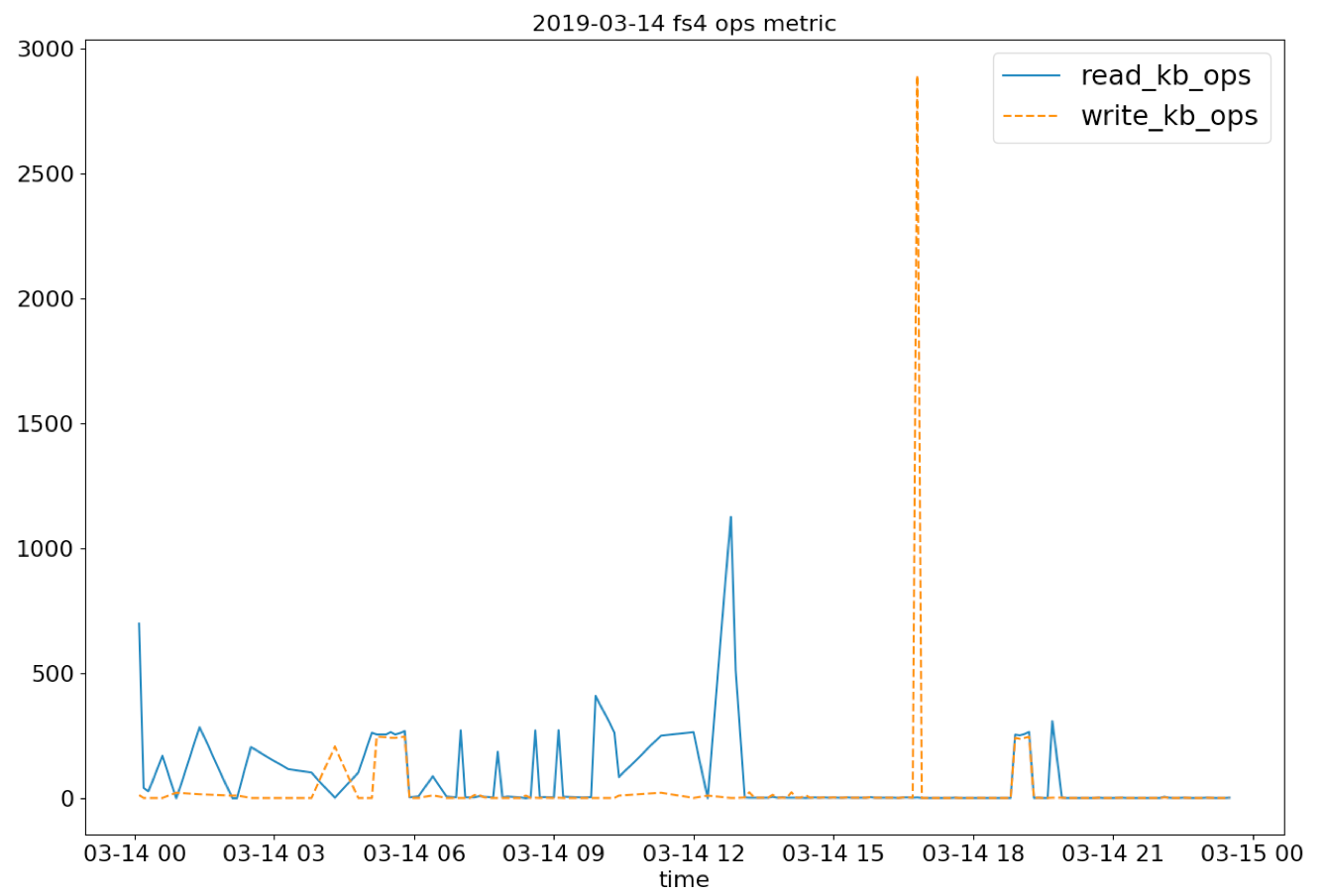}}
\caption{Sample report showing the read and write quality 
to file system fs2 over 24 hours.}
\label{fig-lassi-ops}
\end{figure}

A workflow has been established where Lustre statistics (collected in the LAPCAT database) and application data (from PBS) are exported 
and ingested by LASSi. Daily risk plots are generated and are available to helpdesk staff. LASSi uses Spark~\cite{Zaharia:2016:ASU:3013530.2934664} for data analytics and matplotlib for generating reports. Custom risk plots and raw Lustre operation data plots can also be generated manually.
Figure~\ref{fig-lassi-risk} shows the risk metrics for file system fs2 over a sample period of 24 hours.  
The \emph{oss\_risk} relates to actual data movement operations and the \emph{mds\_risk} to metadata operations, note the significant peak in the evening.

Figure~\ref{fig-lassi-oss-risk}  shows an example of the \emph{oss\_risk} metric over 24 hours attributed to the jobs that were running.  
These plots allow us to focus on particular applications.
We have noticed a particular class of applications that can be problematic: \emph{task farms} as is illustrated from 
Figure~\ref{fig-lassi-mds-risk}.  Each individual application contributes to a significant metadata operation load from the whole job.

We have also found the read and write quality metrics to be useful, an example plot of this metric for fs2 over 24 hours is shown in Figure~\ref{fig-lassi-ops}.  The reason this is important is that small reads or writes to Lustre can keep the file system busy for (presumably) little benefit. 

\begin{figure}[htbp]
\centerline{\includegraphics[scale=0.3]{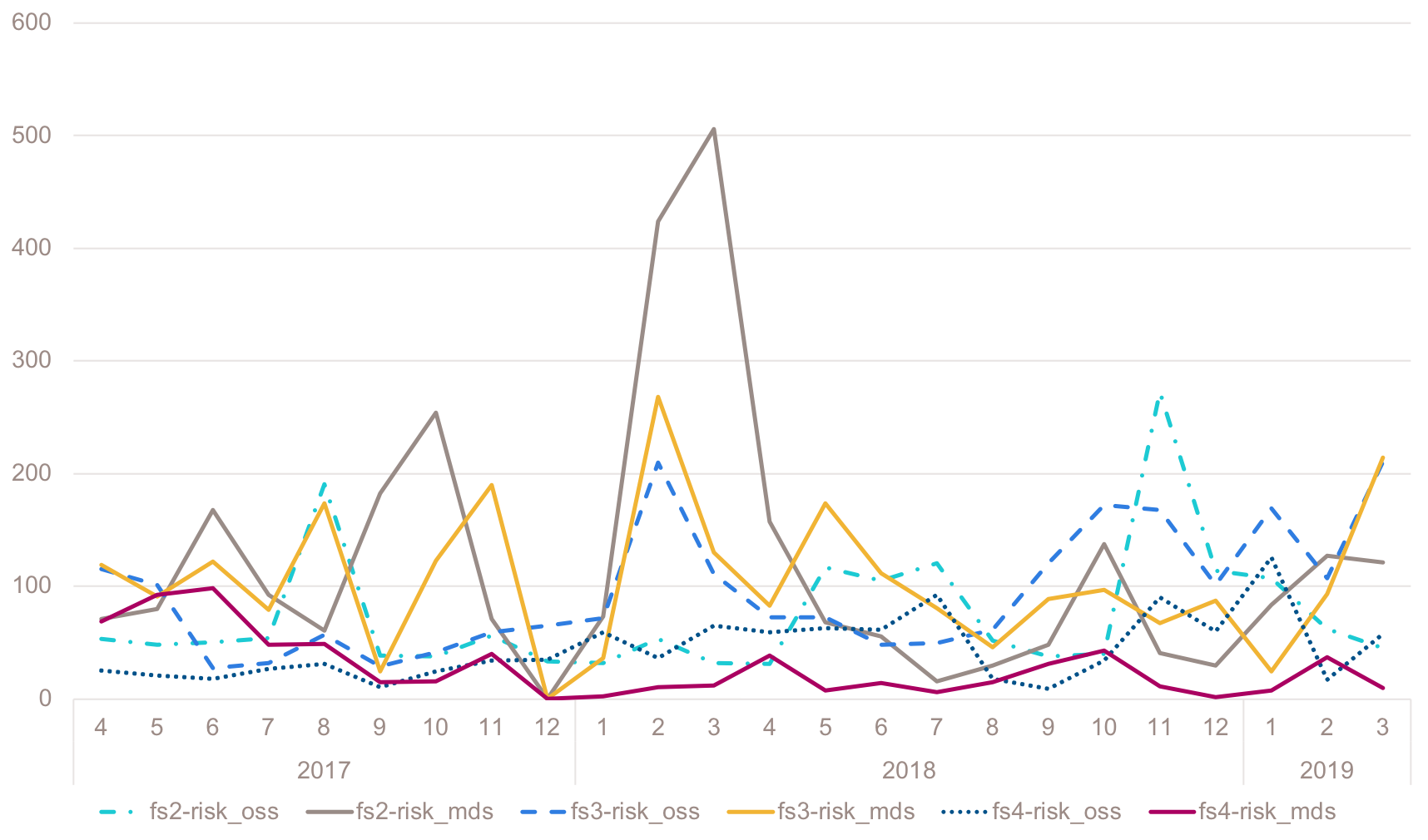}}
\caption{Risk metric of file system averaged over months.}
\label{fig-lassi-risk-fs}
\end{figure}

\begin{figure}[tbp]
\centerline{\includegraphics[scale=0.3]{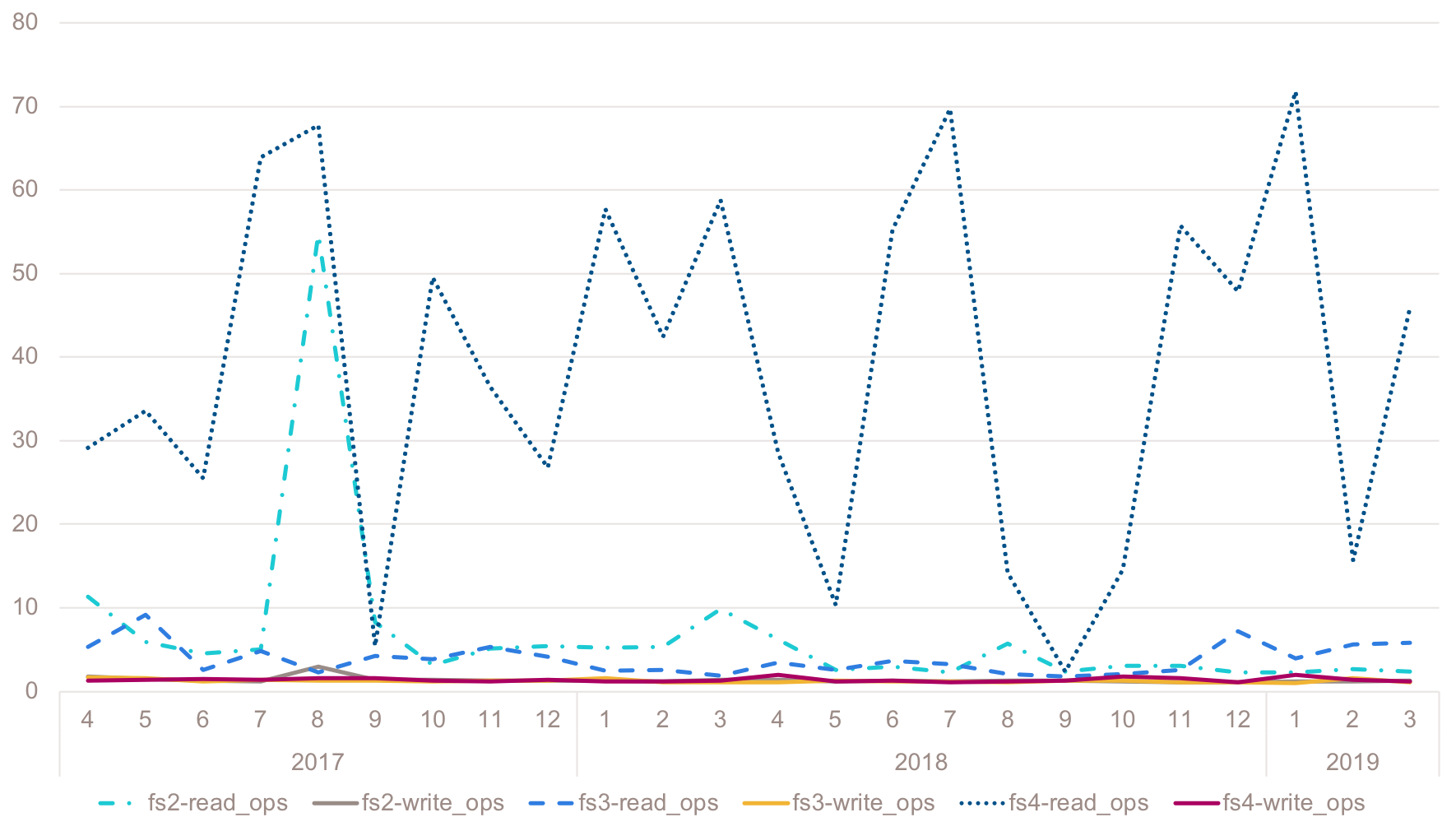}}
\caption{Ops metric of file systems averaged over months.}
\label{fig-lassi-ops-fs}
\end{figure}

Figure~\ref{fig-lassi-risk-fs} shows the variation in overall risk metric over many months and clearly there is a variation in workload during this time with a peak in March for fs2. 
We observe that fs2 and fs3 generally have higher risk than fs4.  For the same period, we show the quality metrics (Figure ~\ref{fig-lassi-ops-fs}) and we can see that reads on fs4 are generally of low quality. 
This file system has the most disparate workload and paradoxically we receive very few complaints over performance in this file system so it is likely that the user base are not heavily dependent on the file system performance.

\subsection{SAFE}

SAFE is an integrated service administration and reporting tool developed and maintained by EPCC~\cite{SAFE}. For this work, it is important to note that SAFE is able to take data feeds from a wide variety of sources and link them in such a way that enables reporting across different system aspects.

We have developed a data feed from LASSi into SAFE that provides the following \textit{aggregated} I/O metrics on a \textit{per job} basis for every job that is run on the ARCHER system:

\begin{itemize}
    \item Total amount of data read.
    \item Total amount of data written.
    \item Total number of read operations.
    \item Total number of write operations.
\end{itemize}

Once ingested into SAFE, these records can then be linked to any other aspects of the job to enable different reporting queries to be performed. For example, we can summarise the I/O data  based on all jobs that belong to a particular research area (by linking with project metadata linked to the job) or we can report on I/O associated with a particular application (by linking with application metadata provided by the Cray Resource Usage Reporting data feed). We have used the first of these linkages in the analysis presented below.

We measure the amount of data written and read by each job in GiB and use this value, along with the job size and the amount of core hours (core-h) spent in the job to compute a two-dimensional heatmap that reveals in which categories of job size and data size the most ARCHER resource is spent. The core-h correspond directly to cost on ARCHER and so using this value as the weighting factor for the heatmaps allows us to assess the relative importance of different I/O patterns.

\subsection{IO-500}

The IO-500\footnote{\url{https://github.com/vi4io/io-500-dev}} is a benchmark suite that establishes I/O performance expectations for naive and optimised access; a single score is derived from the individual measurements and released publicly in a list to foster the competition. Similarly to Top500, a list is released on each ISC-HPC and Supercomputing conference \cite{io500}.

The design goals for the benchmark were: representative, understandable, scalable, portable, inclusive, lightweight, and trustworthy.
The IO-500 is built on the standard benchmarks MDTest and IOR\footnote{\url{https://github.com/hpc/ior}}. The workloads represent:
\begin{itemize}
 \item IOREasy: Applications with well optimized I/O patterns.
 \item IORHard: Applications that require a random workload.
 \item MDEasy: Metadata and small object access in balanced directories.
 \item MDHard: Small data access (3901 bytes) of a shared directory.
 \item Find: Locating objects based on name, size, and timestamp.
\end{itemize}
The workloads are executed in a script that first performs all write phases and then the read phases to minimise cache reuse.

\paragraph{Performance Probing}

To understand the response times for the IO-500 case further, we run a probe every second on a node that measures the response time for accessing a random 1\,MB of data in a 
200\,GB file and for a create, stat, read, delete of one file in a pool of 200k files. 
The I/O test uses the \texttt{dd} tool for access while the metadata test uses MDWorkbench\cite{UMLWMKM19} which allows for such regression testing.
The investigation of the response times enables a fine-grained investigation of the system behavior and to assess the observed risk.

\section{Results and Analysis}

\subsection{LASSi Application Analysis}

In this section we show recent analysis of the application I/O on ARCHER for the period 
April 2017 to March 2019 inclusive (\textit{i.e.} two full years) by characterising them with the \emph{risk} and \emph{ops} metrics. 

\subsubsection{Applications Slowdown Analysis}
LASSi was originally developed to analyse events of slowdown, reported by users.
In the case of a slowdown event, the time window of the event is mapped to the 
file system risk and ops profile. 
This will easily tell us if I/O is responsible
for slowdown and which application was causing the slowdown.
LASSi has historical run time data of all application runs and user reports of application slowdown is always validated to check for actual slowdown.

High \emph{risk\_oss} usually corresponds to a more than average quantity of reads and writes. 
This is generally not concerning since the shared file systems are configured to deliver 
high I/O bandwidth.
In such cases, attention should be given more to the I/O quality as denoted by ops
metric. In case of high MDS risk, the application
should be carefully studied for high metadata operations that contribute to the risk.

In LASSi, applications are grouped by the exact run time command used. Usually
a user reports jobs that ran normally and which ran slower. Sometimes this 
detailed information is not provided. In such cases, LASSi analysis will consider 
all jobs in the group for analysis. Slowdown is a function of the I/O profile of 
the application and the risk and ops profile of the file system that the application
encounters. For instance, an application that does not perform I/O will not be impacted 
by the risk in the file system. Similarly, application with high metadata operations
will be impacted by the $risk\_mds$ and not $risk\_oss$.

This slowdown analysis used to take around a day or two and LASSi has made this
process simple and such analysis are usually done in minutes using the automated 
daily reports. Further development is in progress to automatically identify
application slowdown and identify the causes.

\subsubsection{Applications Usage Analysis}

A useful way to view the risk to the file system from a mix of applications is a scatter plot showing \emph{OSS} and \emph{MDS} risk for a set of applications.
Using the scatter plots, we can identify general trends in file system usage and identify
main issues or usage patterns.
This study of the profile of the \emph{risk} and 
\emph{ops} metrics across file system over a long period is helpful for system architects and service staff to improve operational quality and 
plan for future. Even though we can characterize different file systems based on the metrics, there is usually not a 
strict direct mapping from applications to file system. A more interesting analysis is to study the metrics of each application group. 
In this section we will look at the \emph{risk} and \emph{ops} profile of application groups based on their run command.

\begin{figure}[tbp]
\centerline{\includegraphics[scale=0.39]{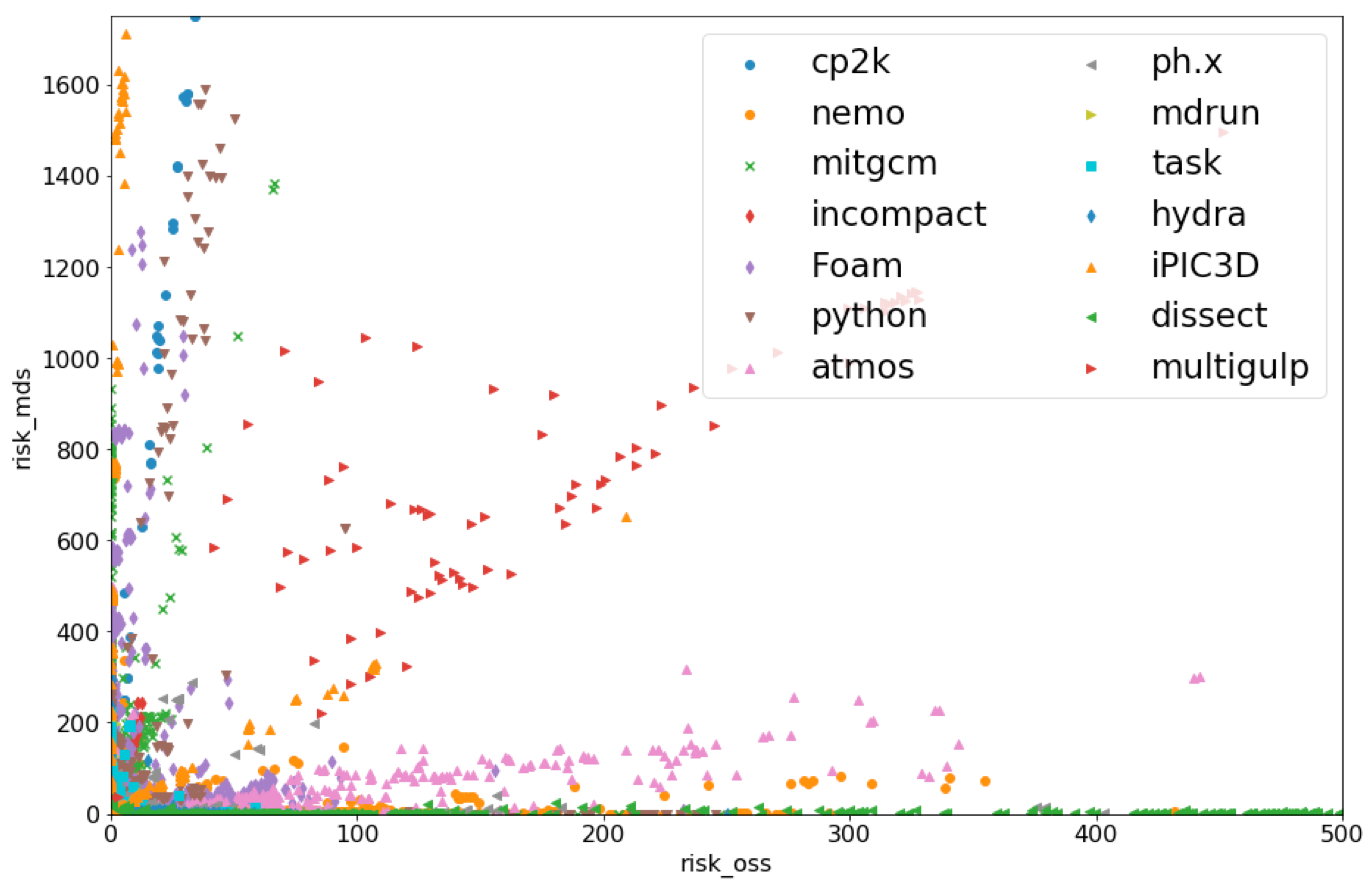}}
\caption{Scatter plot of \emph{risk\_oss} vs \emph{risk\_mds} for applications.}
\label{fig-lassi-risk-apps-low}
\end{figure}

\begin{figure}[tbp]
\centerline{\includegraphics[scale=0.39]{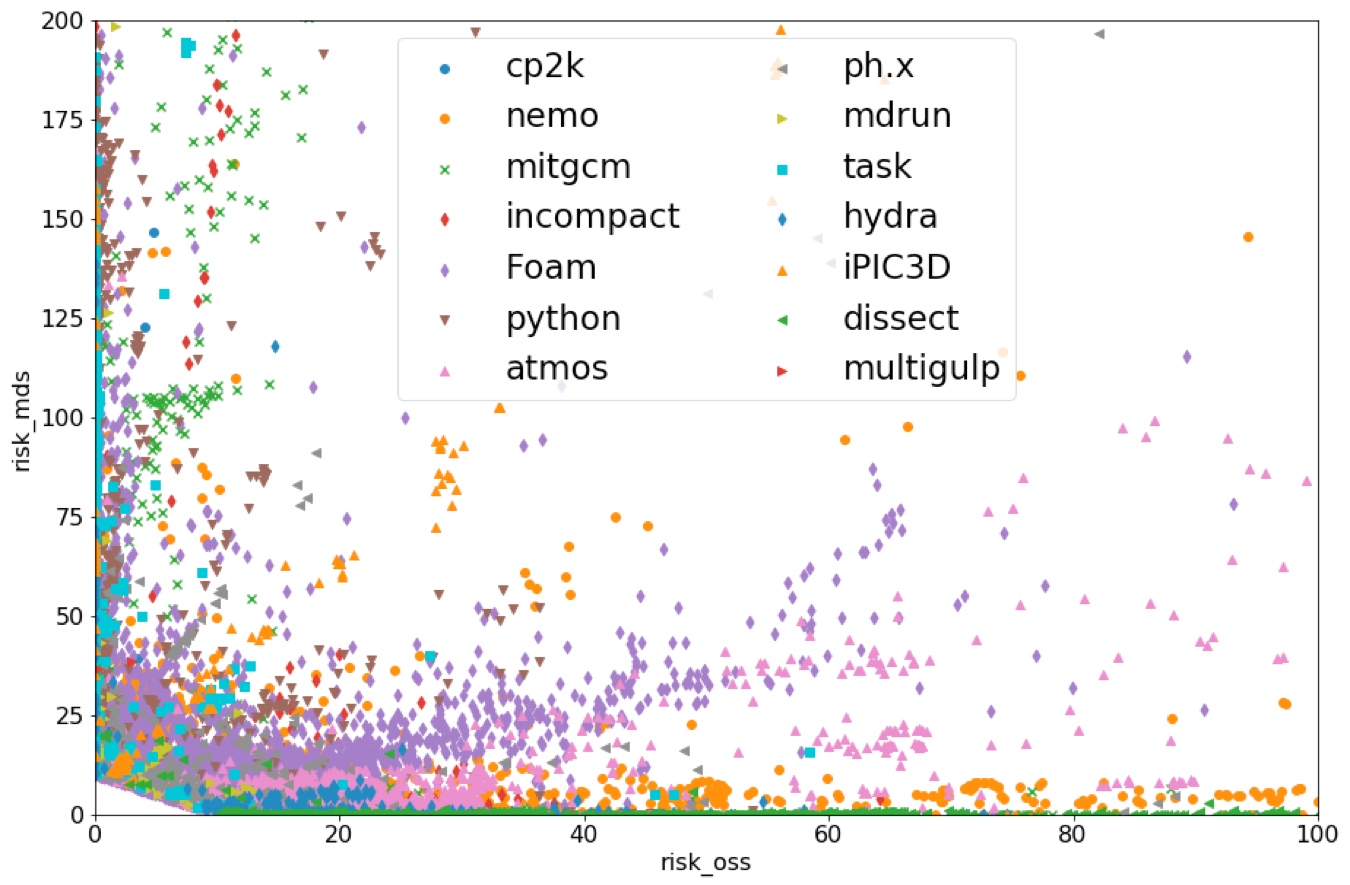}}
\caption{Scatter plot of \emph{risk\_oss} vs \emph{risk\_mds} for applications at high resolution.}
\label{fig-lassi-risk-apps-high}
\end{figure}

We use previous experience gained by the site support team, to map the run command to the application being run.
Figure \ref{fig-lassi-risk-apps-low} shows the scatter plot of
\emph{risk\_oss} vs \emph{risk\_mds} for different application groups. Figure \ref{fig-lassi-risk-apps-high} shows the
same metric for applications zoomed in to the bottom-left corner. For simplicity, 14 application groups are 
shown and we ignore applications with $(risk\_oss + risk\_mds) < 25$. 
The $risk\_oss$ and  $risk\_mds$ in the plots refer to the average value of any application run over its run time.

The first thing to note from Figure~\ref{fig-lassi-risk-apps-low} is a  
pattern of 
risks mostly clustered around the axis for most applications except \emph{multigulp}. The points scattered around the $risk\_oss$ indicates application doing more reads and write using lesser metadata operations.
\emph{dissect}, \emph{atmos} and \emph{nemo} applications follow this pattern
Similarly, the points scattered around the $risk\_mds$ indicates application 
using more metadata operations to complete lesser quantity of reads or writes. This pattern is seen in \emph{iPIC3D}, \emph{Foam}, \emph{cp2k}, \emph{python} and \emph{mitgcm} applications.

The zoomed-in view (in \Cref{fig-lassi-risk-apps-high}) shows a similar pattern of risks mostly clustered around the axis. 
We can see clustering of \emph{hydra} near both the $risk\_oss$ and $risk\_mds$ axis. \emph{incompact} and few instances of \emph{mdrun} application clustered near the $risk\_mds$ axis.
The \emph{ph.x} application show no clear pattern but have many runs with considerable $risk\_oss$ and $risk\_mds$ like the \emph{multigulp} applications. 
There are many instances of \emph{task-farm} like applications that have smaller risk. The risks from \emph{task-farm} get amplified as individual tasks are scheduled to run
in huge numbers at the same time.

\subsubsection{Application profile}

In this section, we will take a more in depth look at the detailed risk and ops profile of four application groups.
Figures \ref{fig-lassi-risk-apps-low} and \ref{fig-lassi-risk-apps-high} show the risk profile of multiple application groups but does not include the I/O quality (ops profile).

\begin{figure}[tbp]
\centerline{\includegraphics[scale=0.39]{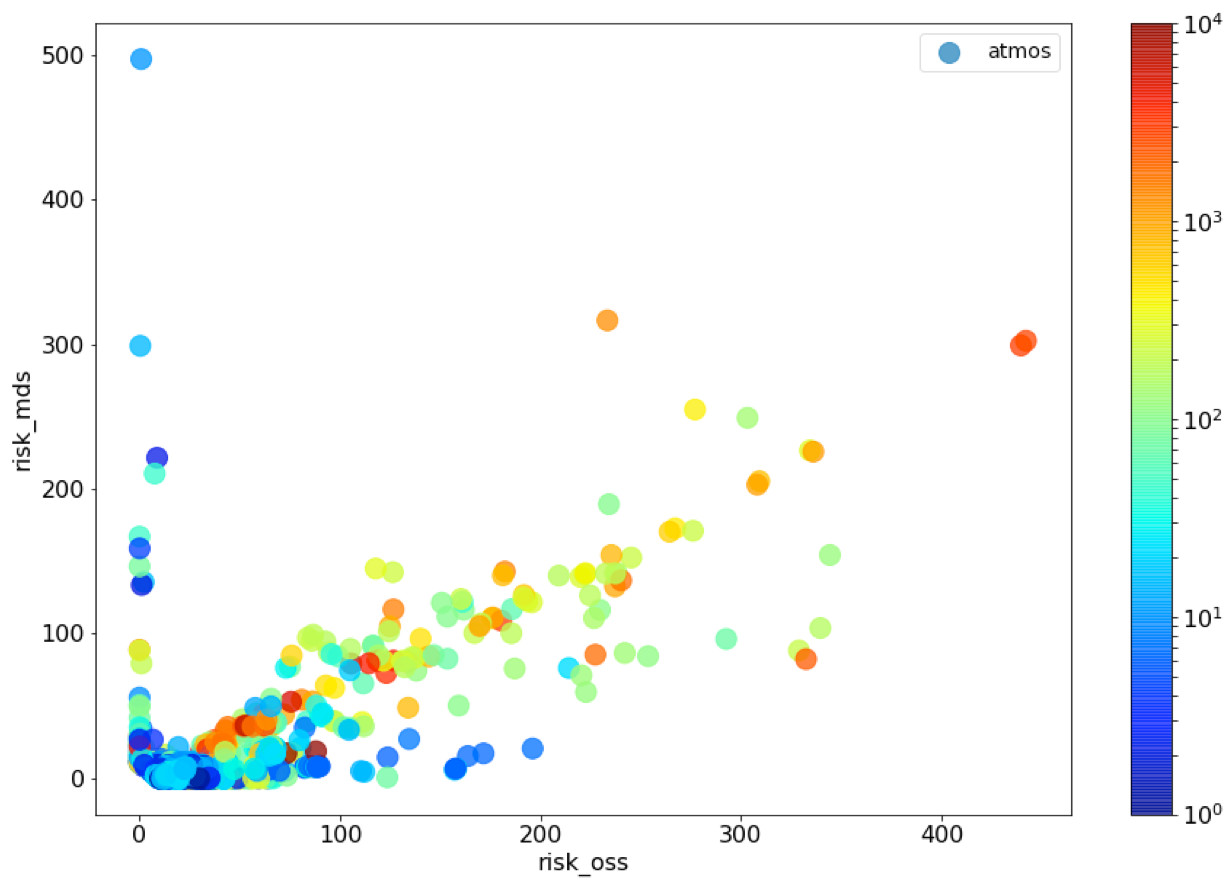}}
\caption{Scatter plot of \emph{risk\_oss} vs \emph{risk\_mds} for Atmos, with color map indicating the I/O quality ($read\_kb\_ops$ + $write\_kb\_ops$).}
\label{fig-lassi-risk-atmos}
\end{figure}

\begin{figure}[htbp]
\centerline{\includegraphics[scale=0.39]{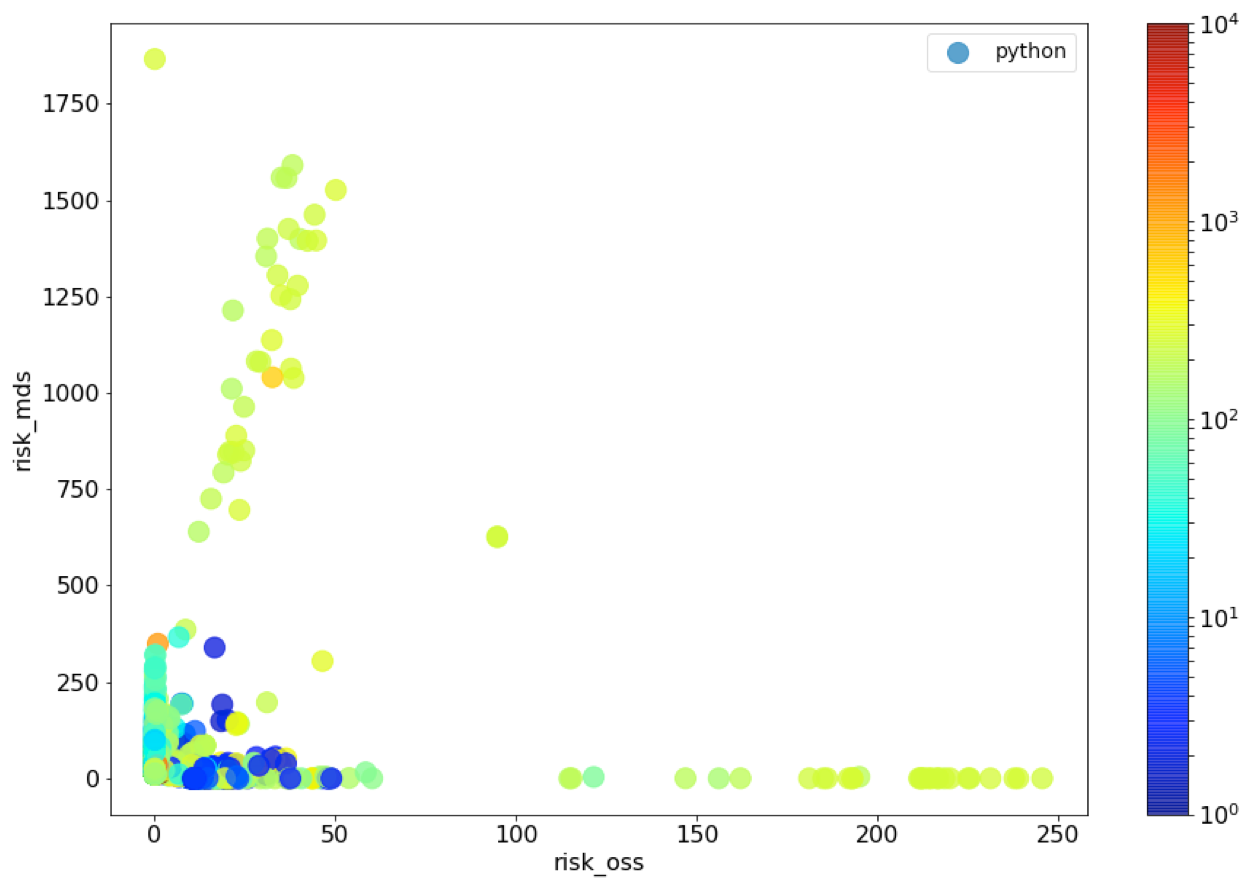}}
\caption{Scatter plot of \emph{risk\_oss} vs \emph{risk\_mds} for Python, with color map indicating the I/O quality ($read\_kb\_ops$ + $write\_kb\_ops$).}
\label{fig-lassi-risk-python}
\end{figure}

\begin{figure}[htbp]
\centerline{\includegraphics[scale=0.39]{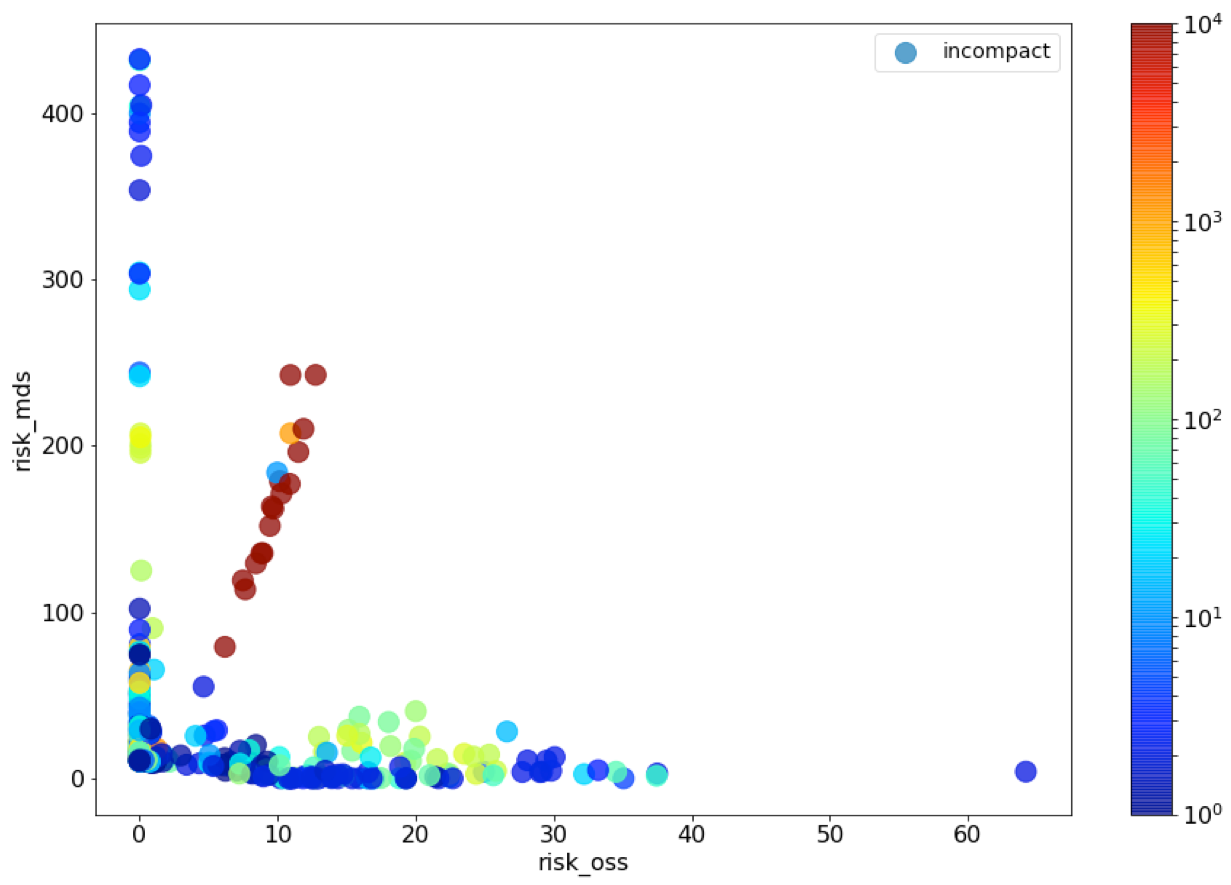}}
\caption{Scatter plot of \emph{risk\_oss} vs \emph{risk\_mds} for Incompact, with color map indicating the I/O quality ($read\_kb\_ops$ + $write\_kb\_ops$).}
\label{fig-lassi-risk-incompact}
\end{figure}

\begin{figure}[htbp]
\centerline{\includegraphics[scale=0.39]{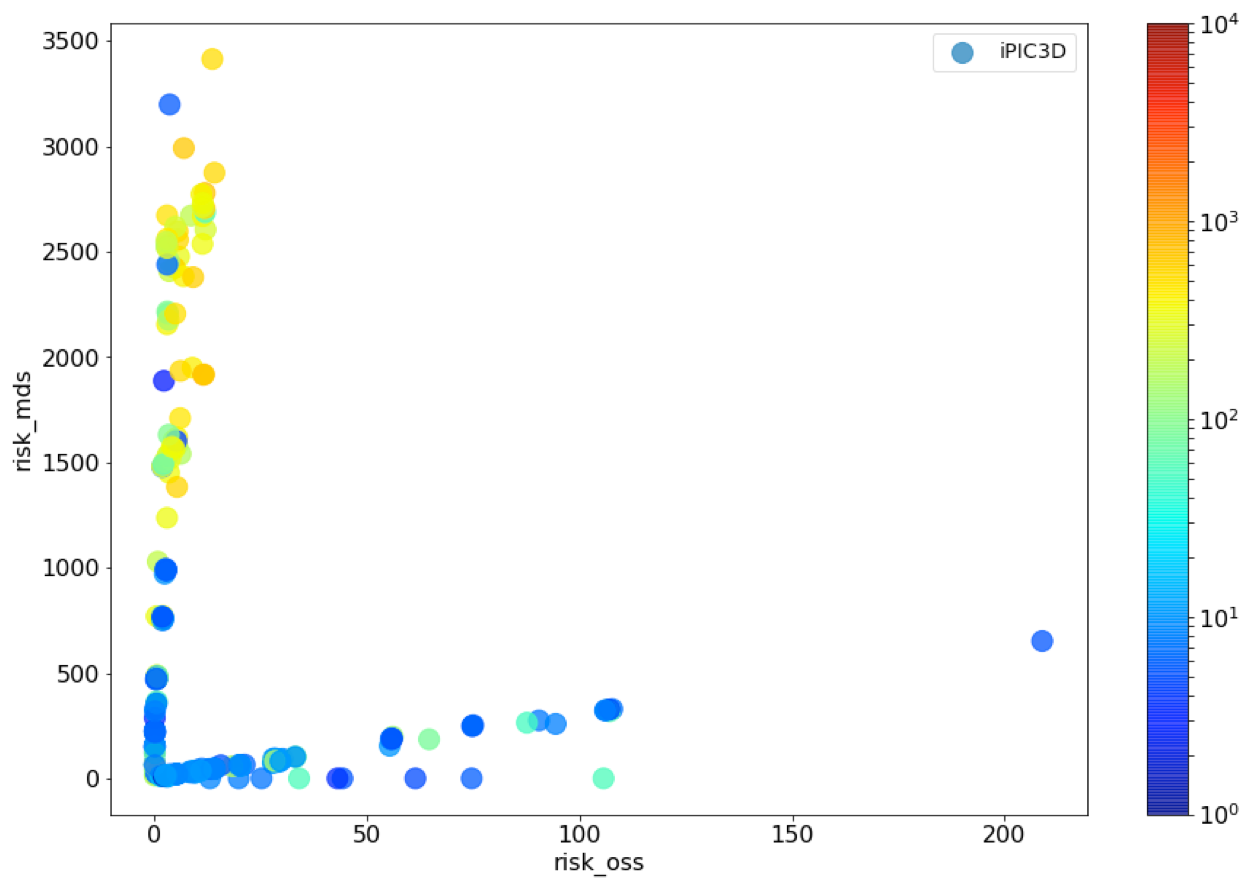}}
\caption{Scatter plot of \emph{risk\_oss} vs \emph{risk\_mds} for \emph{iPIC3D}, with color map indicating the I/O quality ($read\_kb\_ops$ + $write\_kb\_ops$).}
\label{fig-lassi-risk-ipic3d}
\end{figure}

Figures \ref{fig-lassi-risk-atmos}, \ref{fig-lassi-risk-python}, \ref{fig-lassi-risk-incompact} and \ref{fig-lassi-risk-ipic3d} show the risk and ops profile of the \emph{atmos}, \emph{python}, \emph{incompact} and \emph{iPIC3D} applications respectively.
All plots show scatter of $risk\_oss$ vs $risk\_mds$, with the color map showing the I/O quality ($read\_kb\_ops$ + $write\_kb\_ops$). 
Blue denotes best I/O quality and red, worse I/O quality.

\begin{figure*}[p]
    \centering
    \subcaptionbox{Risk}{
    \hspace*{1mm}
    \includegraphics[width=0.985\textwidth]{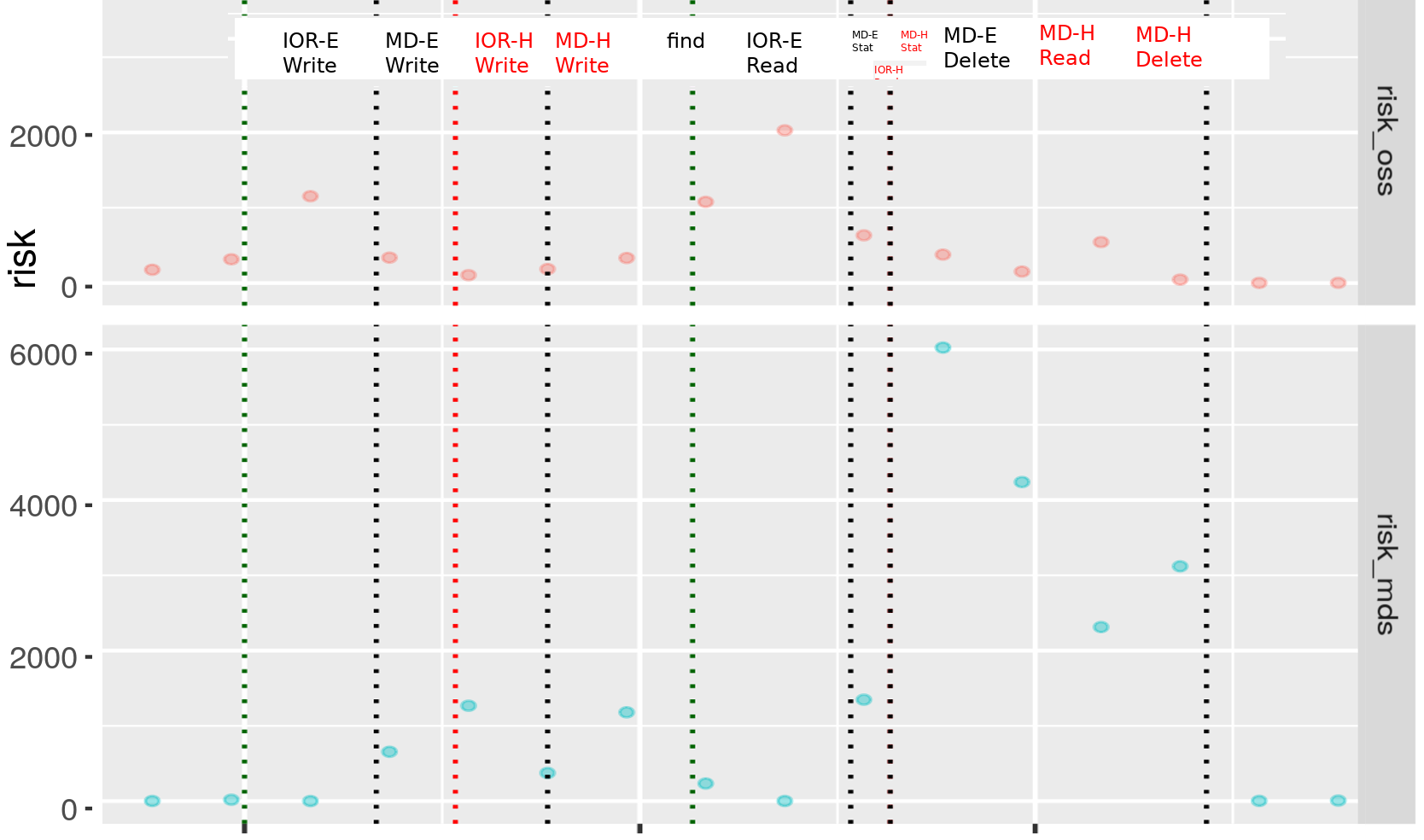}
    }

    \subcaptionbox{Response time as measured by the probing}{\includegraphics[width=\textwidth]{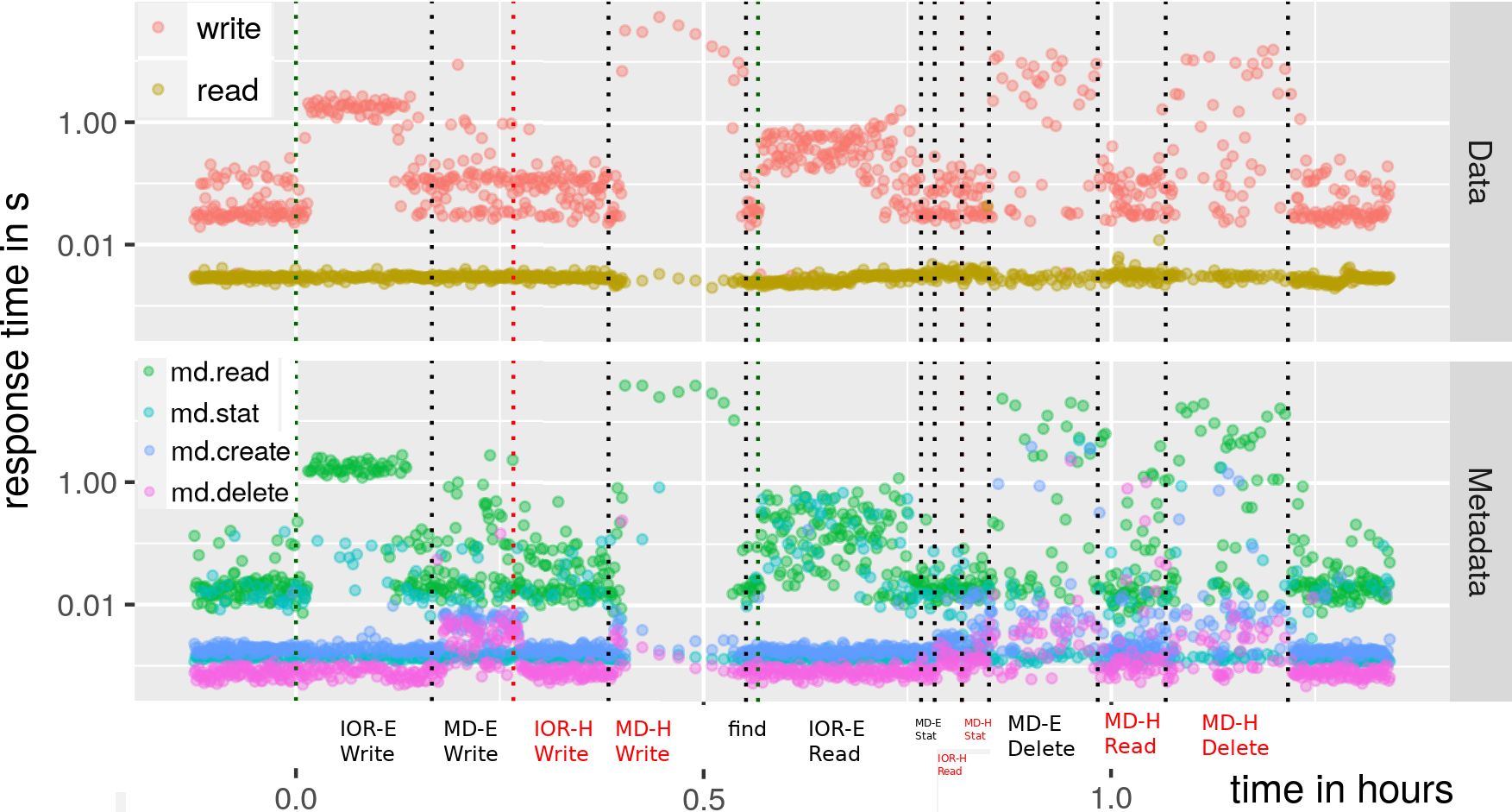}}%

    \caption{Observed behavior of the IO-500 on 100 ARCHER nodes.}
    \label{fig:io500}
\end{figure*}

The clusters in Figure \ref{fig-lassi-risk-atmos}, the \emph{atmos} applications reveal three different I/O patterns. 
Clusters near the axis show good I/O quality whereas the cluster away from the axis shows poor I/O quality. 
Clusters of \emph{python} applications in Figure \ref{fig-lassi-risk-python}, show both high metadata and \emph{OSS} usage, 
but in general suffer from poor I/O quality, whereas some application with low risk perform good I/O quality.    

Most \emph{incompact} applications in Figure \ref{fig-lassi-risk-incompact} show good I/O quality whereas a cluster of application runs away from the axis show very bad I/O quality. 
Many \emph{iPIC3D} application are characterised by high metadata usage and bad I/O quality as shown in Figure \ref{fig-lassi-risk-incompact}. A cluster of \emph{iPIC3D} application with high OSS risk have good I/O quality.

We see a general trend in application profiles that there is variance in both the quantity and I/O quality but they all show clear trends as seen by the clustering.
This clearly points to different application configurations used by researchers. 
It is encouraging to see many application runs showing good I/O quality and high amounts of I/O. Understanding why different application runs in the same scientific community have lower I/O quality or use more metadata operations is important and we plan to investigate this further in the future.

\subsection{IO-500 Probes and LASSi}

To investigate the behavior of the risk for running applications, we executed the IO-500 benchmark on 100 nodes on ARCHER. 
The benchmark reported for the different phases the following performance values: (IOREasy\_write: 12.973 GB/s,
MDEasy\_write: 58.312 kiops,
IORHard\_write: 0.046 GB/s,
MDHard\_write: 34.324 kiops,
find: 239.300 kiops,
IOREasy\_read: 9.823 GB/s,
MDEasy\_stat: 64.173 kiops,
IORHard\_read: 1.880 GB/s,
MDHard\_stat: 63.166 kiops,
MDEasy\_delete: 13.195 kiops,
MDHard\_read: 20.222 kiops,
MDHard\_delete: 10.582 kiops) with a total IO-500 score of 8.45.

The observed risk is shown in \Cref{fig:io500}(a).
Be aware that due to the reporting interval, the data points cover the 6 minute period left of them (\textit{i.e.} the previous 6 minutes).
We can see that the OSS risk is high during the IOR easy phases, reaching 2000 for the read phase.
The value is around 500 during the MDHard Read phase. 
The IOHard values cannot be recognized from the OSS risk.

Looking at the metadata risk, the MD workloads can be identified; high peaks are seen in the hard workloads towards the end.

To understand the impact on the user perspective, we also run the periodic probing and reported the response time in \Cref{fig:io500}(b) for metadata rates and I/O. 
The data response time correlates well with the risk for IOREasy patterns, the response times are high compared to the risk for the MD hard write and MD delete.
The metadata risk and the metadata shows some correlation particularly to md.delete, but small I/O (md.read) is also delayed significantly for some patterns.

This analysis gives us confidence that the LASSi risk metrics correspond
to real, observable effects on the file systems studied.

\subsection{SAFE Analysis of LASSi Data}

For the SAFE analysis of LASSi data we considered all jobs that ran
on ARCHER in the 6-month period July to December 2018.

\subsubsection{Overall view}

\begin{figure*}[tb]
\centering
\includegraphics[width=\textwidth]{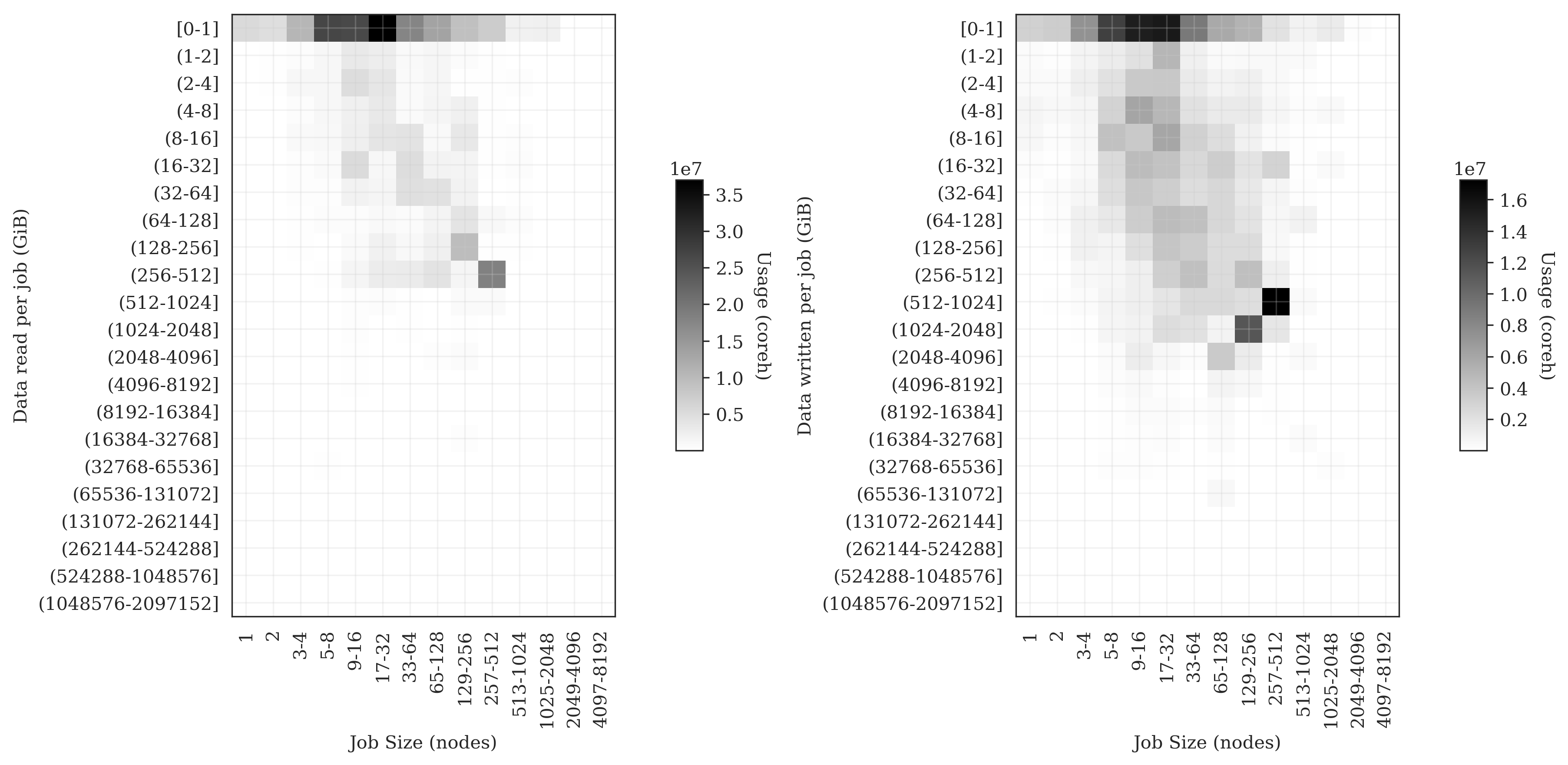}
\caption{Heatmaps of data read per job and data written per job vs job size. Weights correspond to total core-h spent in a particular category.}
\label{fig:safe_data_all}
\end{figure*}

\begin{figure*}[tb]
\centering
\includegraphics[width=\textwidth]{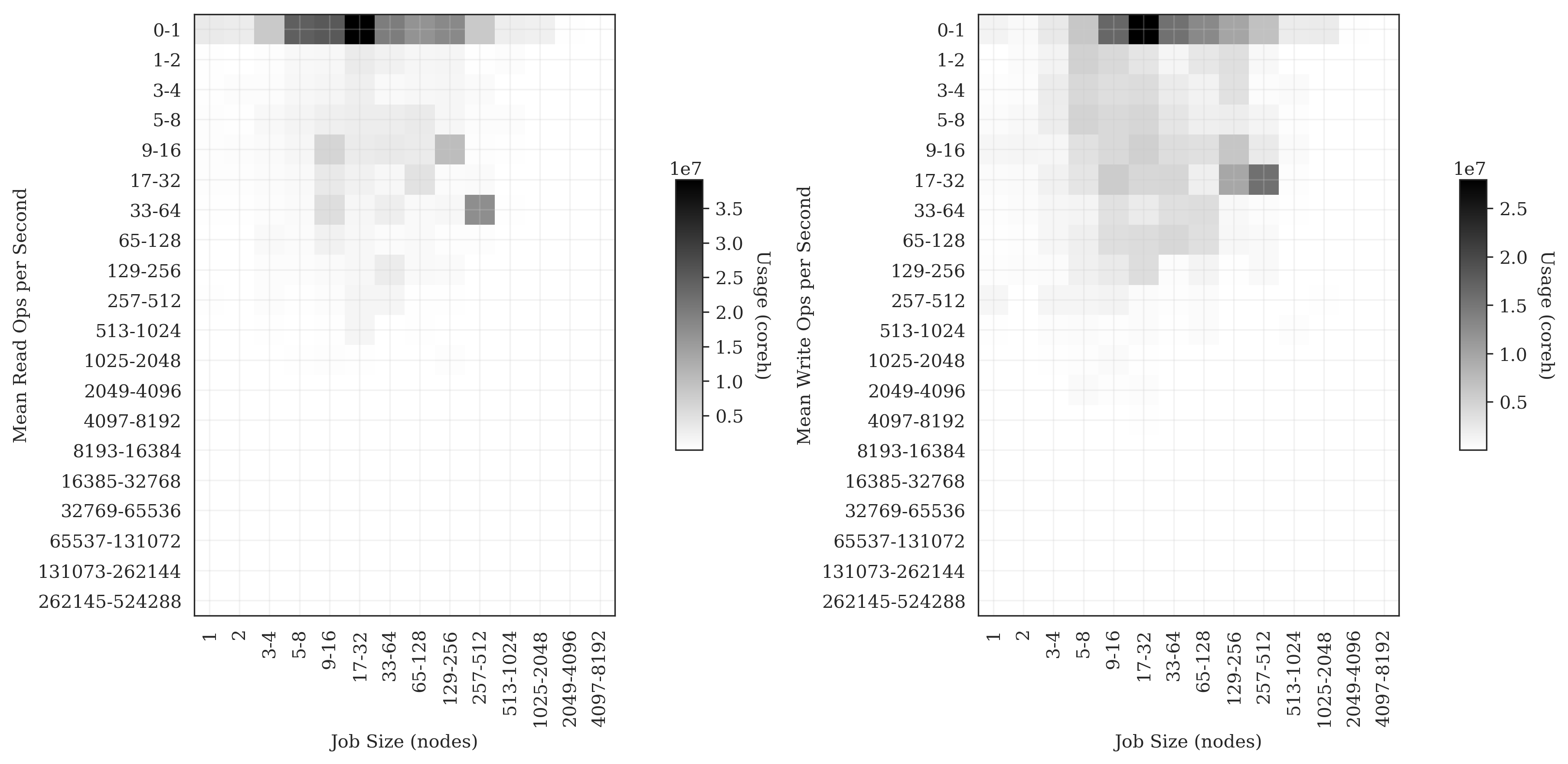}
\caption{Heatmaps of mean read ops/s per job and mean write ops/s per job  vs job size. Weights correspond to total core-h spent in a particular category.}
\label{fig:safe_ops_all}
\end{figure*}

Figures~\ref{fig:safe_data_all} and \ref{fig:safe_ops_all} show I/O heatmaps
for data read, data written, mean read ops/s and mean write ops/s for all jobs
on ARCHER during the analysis period (Jul-Dec 2018 inclusive).

Table~\ref{tab:safe_data_all} summarises the percentage use by amount of data read or written per job for the same period.

In total, 11,279.4~TiB of data were read and 22,094.3~TiB of data were read by 
all jobs on ARCHER during the six month analysis period.

\begin{table}[tbh]
\centering
\caption{\% usage breakdown by data read and written for all jobs run on ARCHER during the analysis period.}
\label{tab:safe_data_all}
\begin{tabular}{ r r r } 
\hline
\hline
Total data per job & \multicolumn{2}{c}{Usage} \\
(GiB) & Read  & Write \\ 
\hline
(0,4) & 59.8\% & 34.8\%\\ 
{[}4,\,32) & 14.7\% & 21.5\%\\ 
{[}32,\,256) & 13.4\%  & 17.8\%\\ 
{[}256,\,2048) & 11.1\%  & 21.4\%\\ 
{[}2048,) & 1.0\%  & 4.5\%\\ 
\hline
\hline
\end{tabular}
\end{table}

The table and heatmaps reveal that a large amount of resources are consumed by jobs that do not read or write large amounts of data (less than 4 GiB read/written per job). We can also see that there are large amounts of use in some categories with large amounts of data written per job - particularly at 129-256 nodes with 1-2 TiB written per job and 257-512 nodes with 0.5-1 TiB written per job. There is a broad range of use writing from 2 to 512 GiB per job in the job size range from 8 to 512 nodes. We note that the analysis shows that user jobs on ARCHER generally read less data than they write by roughly a factor of two.

Figure~\ref{fig:safe_ops_all} heatmaps of I/O operations provide less useful information. As the data ingested into SAFE only contains the total number of operations over the whole job, the computed mean I/O rate is generally small and we would expect that it is the peak rate (in terms of operations per second) that would be required to provide additional insight. For this reason, we constrain our remaining analysis of the LASSi data in SAFE to the total amounts of data read and written per job. We do plan, in the future, to import the peak ops/s rate into SAFE to facilitate useful analysis of this aspect of I/O.

\begin{figure*}[tb]
\centering
\includegraphics[width=\textwidth]{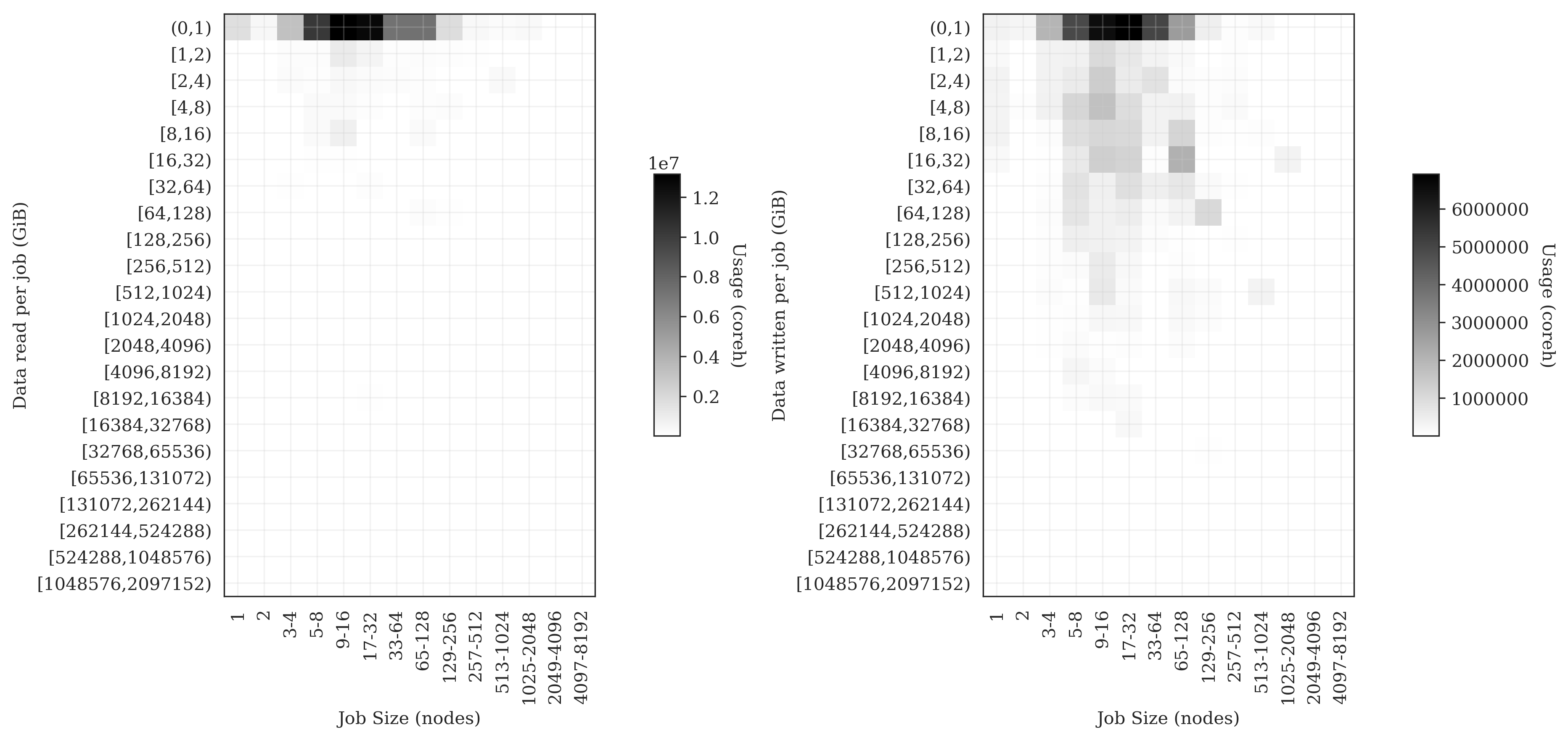}
\caption{Heatmaps of data read per job and data written per job vs job size for the materials science community. Weights correspond to total core-h spent in a particular category.}
\label{fig:safe_data_e05}
\end{figure*}
\begin{figure*}[tb]
\centering
\includegraphics[width=\textwidth]{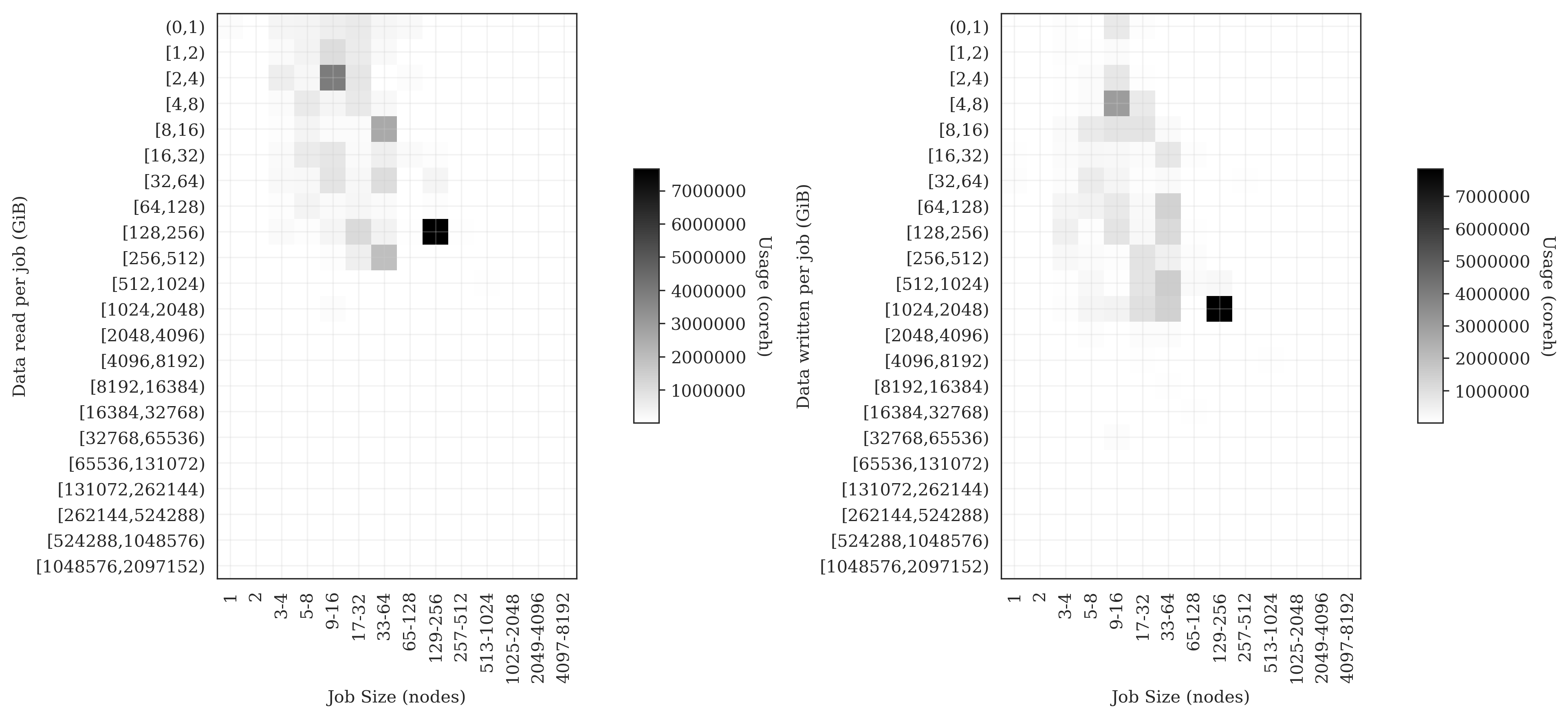}
\caption{Heatmaps of data read per job and data written per job vs job size for the climate modelling community. Weights correspond to total core-h spent in a particular category.}
\label{fig:safe_data_n02}
\end{figure*}
\begin{figure*}[tb]
\centering
\includegraphics[width=\textwidth]{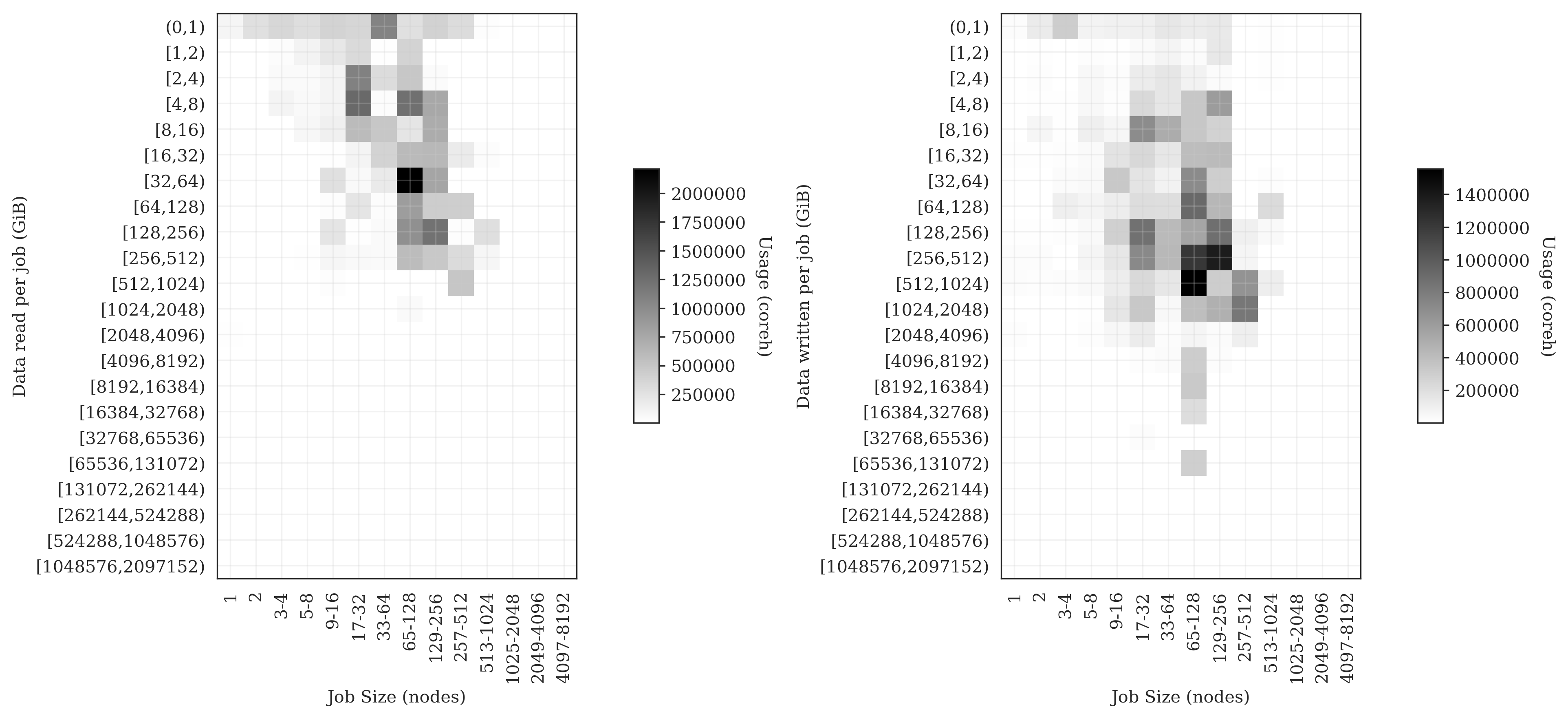}
\caption{Heatmaps of data read per job and data written per job vs job size for the CFD community. Weights correspond to total core-h spent in a particular category.}
\label{fig:safe_data_e01}
\end{figure*}
\begin{figure*}[tb]
\centering
\includegraphics[width=\textwidth]{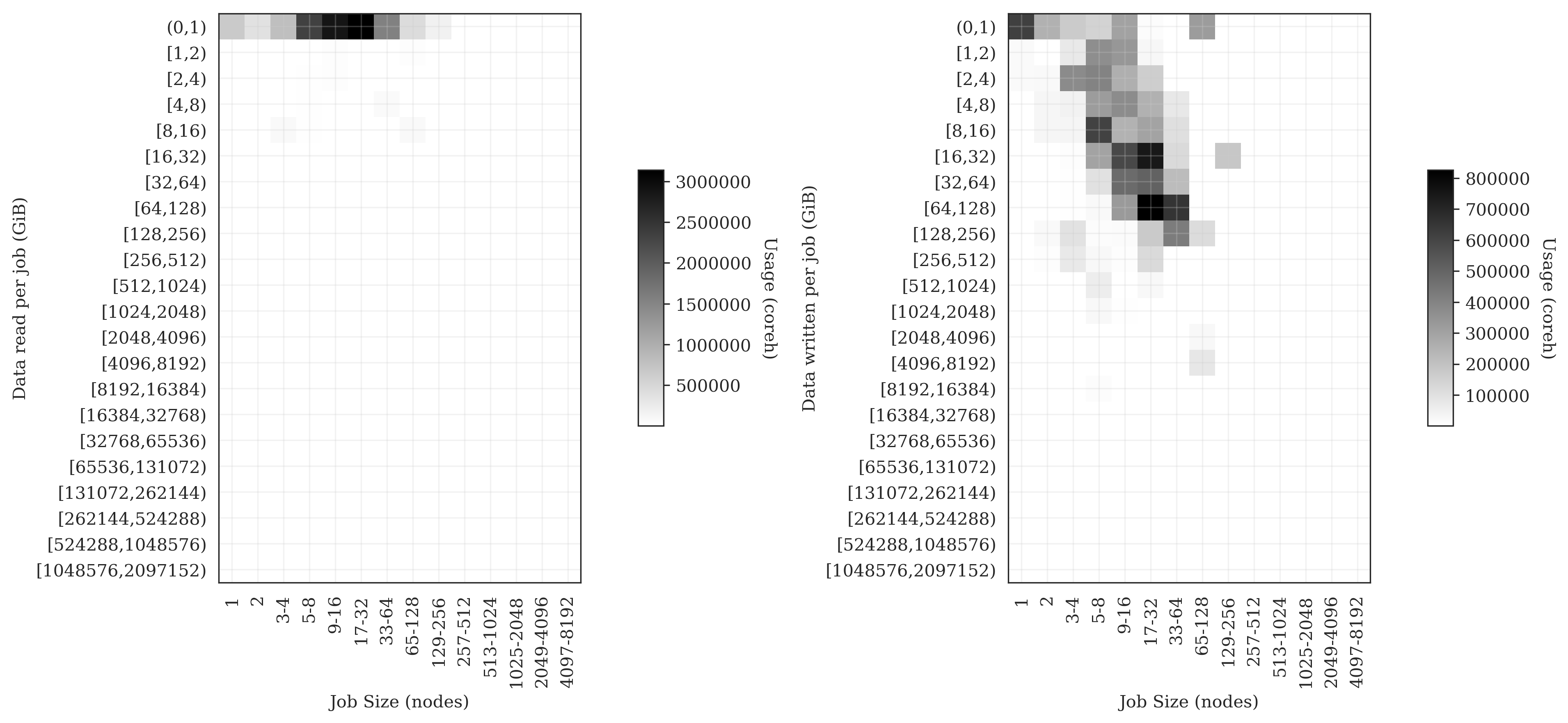}
\caption{Heatmaps of data read per job and data written per job vs job size for the biomolecular modelling community. Weights correspond to total core-h spent in a particular category.}
\label{fig:safe_data_e280}
\end{figure*}

As demonstrated by the LASSi application use analysis, the data for all
jobs within the analysis period will be a overlay
of many different I/O use patterns. In order to start to understand and
identify these different use patterns, the following sections analyse
the I/O patterns for different research communities on ARCHER. In this initial analysis, we consider
four different communities that make up a large proportion of the core hours used on
the service in the analysis period:

\begin{itemize}
    \item Materials science.
    \item Climate modelling.
    \item Computational fluid dynamics (CFD).
    \item Biomolecular modelling.
\end{itemize}

Together, these communities typically account for around 60\% of the total usage on the
ARCHER service. Our initial analysis has focussed on communities with large amounts
of core-h use in the analysis period as core-h use corresponds directly to how resources
are allocated on the service. Future analyses will examine use cases which use large 
amounts of I/O resource without a corresponding large amount of core-h use to allow us to distinguish other I/O use patterns.

\subsubsection{Materials science}

Materials science research on ARCHER is dominated by the use of
periodic electronic structure applications such as VASP, CASTEP, CP2K and
Quantum Espresso. The I/O heatmap for this community can be seen in
Figure~\ref{fig:safe_data_e05} and the breakdown of data read and written
in Table~\ref{tab:safe_data_e05}. 
In the six month analysis period, the materials science community read a
total of 1,219.0~TiB and wrote a total of 3,795.1~TiB. Note that the
total disk quota for this community on the ARCHER Lustre file systems
is 244~TiB so much of the data read/written is transient in some way.

\begin{table}[tbh]
\centering
\caption{Percent usage breakdown by data read and written for all jobs run by
materials science community on ARCHER during the analysis period.}
\label{tab:safe_data_e05}
\begin{tabular}{ r r r } 
\hline
\hline
Total data per job & \multicolumn{2}{c}{Usage} \\
(GiB) & Read  & Write \\ 
\hline
(0,4) & 94.3\% & 55.4\%\\ 
{[}4,\,32) & 4.2\% & 25.0\%\\ 
{[}32,\,256) & 1.1\%  & 12.3\%\\ 
{[}256,\,2048) & 0.4\%  & 5.1\%\\ 
{[}2048,) & 0.2\%  & 2.2\%\\ 
\hline
\hline
\end{tabular}
\end{table}

It is obvious that the vast majority of materials science research on ARCHER
does not have large requirements on reading or writing large amounts of data
\textit{on a per job basis}. However, due to the large amount of use
associated with this community, they still manage to read and write large amounts
of data in total even though the amount per job is small. In most cases, for
the applications used and research problems treated by this community this 
I/O pattern can be understood as:

\begin{itemize}
    \item the input data is small: often just a description of the initial atomic 
coordinates, basis set specification and a small number of calculation parameters;
    \item the output data is also small: including properties of the modelled system
such as energy, final atomic coordinates and descriptions of the wave function.
\end{itemize}

Closer inspection of the data shows that there is significant usage (37.3\%) for
jobs that write larger amounts of data ([4,\,256)~GiB). We
expect these jobs to correspond mostly to cases where users are running 
dynamical simulations where the time trajectories of properties of the system
being modelled are captured for future analysis.

In the future, we expect the size of systems modelled in this community to stay
largely static and so the I/O requirement for individual jobs will not
increase significantly. However, the drive to more statistically-demanding
sampling of parameter space in this community will drive an overall increase
in I/O requirements going forwards.

\subsubsection{Climate modelling}

This research is dominated by the use of
applications such as the Met Office Unified Model, WRF, NEMO and
MITgcm. The I/O heatmap for this community can be seen in
Figure~\ref{fig:safe_data_n02} and the breakdown of data read and
written in Table~\ref{tab:safe_data_n02}. The climate modelling community 
read a total of 503.5~TiB and wrote a total of 2,404.5~TiB in the six month
analysis period. The disk quota for this community on the ARCHER
Lustre file systems is 541 TiB.

\begin{table}[tbh]
\centering
\caption{Percent usage breakdown by data read and written for all jobs run by
climate modelling community on ARCHER during the analysis period.}
\label{tab:safe_data_n02}
\begin{tabular}{ r r r } 
\hline
\hline
Total data per job & \multicolumn{2}{c}{Usage} \\
(GiB) & Read  & Write \\ 
\hline
(0,4) & 30.0\% & 6.3\%\\ 
{[}4,\,32) & 22.4\% & 24.0\%\\ 
{[}32,\,256) & 39.8\%  & 21.1\%\\ 
{[}256,\,2048) & 7.8\%  & 46.4\%\\ 
{[}2048,) & 0.0\%  & 2.2\%\\ 
\hline
\hline
\end{tabular}
\end{table}

The climate modelling community typically read and write large amounts of 
data per job with the largest use in the per-job read interval [32,\,256)~GiB and
the largest use in the per-job write interval [256,\,2048)~GiB. This pattern
can be understood as:

\begin{itemize}
    \item most jobs read in large amounts of observational data and model
    description data;
    \item most jobs write out time-series trajectories of the model configuration
    and computed properties for a number of snapshots throughout the model run.
    These trajectories are archived and used for further analysis.
\end{itemize}

The size of the output trajectories is intrinsically linked to the resolution
of the model being used for the research and so we would expect the I/O 
requirements of \textit{individual jobs} from this community to increase
as the resolution of models increases.

\subsubsection{Computational fluid dynamics (CFD)}

CFD research on ARCHER is
dominated by the use of applications such as SBLI, OpenFOAM,
Nektar++ and HYDRA. The I/O heatmap for this community can be seen in
Figure~\ref{fig:safe_data_e01} and the breakdown of data read and written
in Table~\ref{tab:safe_data_e01}. The CFD community 
read a total of 205.2 TiB and wrote a total of 1,016.7 TiB in the six month
analysis period. The disk quota for this community on the ARCHER
Lustre file systems is 352 TiB.

\begin{table}[tbh]
\centering
\caption{Percent usage breakdown by data read and written for all jobs run by
CFD community on ARCHER during the analysis period.}
\label{tab:safe_data_e01}
\begin{tabular}{ r r r } 
\hline
\hline
Total data per job & \multicolumn{2}{c}{Usage} \\
(GiB) & Read  & Write \\ 
\hline
(0,4) & 27.6\% & 7.7\%\\ 
{[}4,\,32) & 30.7\% & 19.5\%\\ 
{[}32,\,256) & 32.8\%  & 28.4\%\\ 
{[}256,\,2048) & 8.5\%  & 37.9\%\\ 
{[}2048,) & 0.4\%  & 8.5\%\\ 
\hline
\hline
\end{tabular}
\end{table}

Table~\ref{tab:safe_data_e01} shows a very similar high-level profile to that for
the climate modelling community (Table~\ref{tab:safe_data_n02}) however,
there is a larger difference in the distribution of usage shown in
Figure~\ref{fig:safe_data_e01} when compared to that for the climate modelling
community (Figure~\ref{fig:safe_data_n02}). The high-level similarity
can be understood due to the similarity in technical setup between the
two communities: jobs for both communities use grid-based modelling approaches,
need to read in large model descriptions and write out time-series trajectories
with large amounts of data. The difference in the distribution of use can
be understood due to the wider range of modelling scenarios used within the
CFD community compared to the climate modelling community. Climate models
have a small range of scales (in terms of length and timescale) when compared
to CFD models, where the systems being studied can range in size from the
tiny (e.g. flow in small blood vessels) to the very large (e.g. models of
full offshore wind farms) and also encompass many different orders of 
magnitude of timescales.

Going forwards, we expect the diversity of modelling scenarios to remain for
the general CFD community with, similarly to the climate modelling community,
a corresponding drive to higher resolution in most use cases leading to an
increase in the I/O requirements on a \textit{per job} basis.

\subsubsection{Biomolecular modelling}

Biomolecular modelling research on ARCHER is
dominated by the use of applications such as GROMACS, NAMD and
Amber. The I/O heatmap for this community can be seen in
Figure~\ref{fig:safe_data_e280} and the breakdown of data read
and written in~Table \ref{tab:safe_data_e280}. 
The biomolecular modelling community  read a total of 1.4~TiB
and wrote a total of 197.0~TiB in the six month analysis period.
The disk quota for this community on the ARCHER
Lustre file systems is 26 TiB.

\begin{table}[tbh]
\centering
\caption{Percent usage breakdown by data read and written for all jobs run by
biomolecular modelling community on ARCHER during the analysis period.}
\label{tab:safe_data_e280}
\begin{tabular}{ r r r } 
\hline
\hline
Total data per job & \multicolumn{2}{c}{Usage} \\
(GiB) & Read  & Write \\ 
\hline
(0,4) & 97.9\% & 30.5\%\\ 
{[}4,\,32) & 2.1\% & 34.4\%\\ 
{[}32,\,256) & 0.0\%  & 32.6\%\\ 
{[}256,\,2048) & 0.0\%  & 2.8\%\\ 
{[}2048,) & 0.0\%  & 0.9\%\\ 
\hline
\hline
\end{tabular}
\end{table}

The overall I/O use profile seen for the biomolecular modelling community
differs from those already seen for the other communities investigated: in 
particular, jobs in this community read in small amounts of data (similar
to the materials science community) but write out larger amounts of data
(though not generally as large as the climate modelling and CFD communities
which use grid-based models). In addition, the usage heatmaps reveal that
this community uses smaller individual jobs than the communities using
grid-based models and that the amount of data written is roughly 
correlated with job size. We interpret the I/O use profile in the following
way:

\begin{itemize}
    \item The small amount of data that is read in corresponds to the small
    amount of data required to specify the model system and parameters. In 
    a similar way to jobs in the materials science community, all that is 
    required to describe the model system are initial particle positions and
    a small number of model parameters.
    \item The larger amount of data written when compared to the materials 
    science community is because the majority of jobs produce trajectories 
    with the model system details saved at many snapshots throughout the
    job to be used for further analysis after the job has finished.
\end{itemize}

In the future we do not expect the I/O requirements for individual jobs to
change very much (as the size of biomolecular systems to be studied will not
change dramatically); however, as for the materials science jobs, we expect the
overall I/O requirements to increase as more jobs need to be run to be able to 
perform more complex statistical analyses of the systems being studied.

\section{Summary and Conclusions}

We have outlined our approach to gaining better understanding of how applications on ARCHER interact with the file systems using a combination of the Cray LASSi framework and the
EPCC SAFE software. The LASSi framework takes a risk-based approach to identifying behaviour likely to cause contention in the file systems. This risk based approach has not only been successful in analysing all reported incidents of slowdown but also incidents where a reported slowdown was not related to I/O but had another cause.
LASSi has been used to deliver faster triage of issues and provide a basis for further analysis of how different applications are using the file systems. 

LASSi provides automated daily reports that are available to helpdesk staff.
We demonstrated how LASSi provides holistic I/O analysis by monitoring file system I/O,
generating coarse I/O profile of file systems and application runs along with analysis of
application slowdown using metrics. This application-centric, non-invasive, metric-based
approach has been used successfully in studying application I/O patterns and could be used
for better management of file system and application projects.  We have also shown how a
file system probing approach using IO-500 complements the risk-based approach and validates
it. Here, from the user perspective, the single risk metric provides a good indicator but
does not reflect the observed slowdown in all cases.

SAFE provides a way to combine data and metrics from LASSi with other data feeds
from the ARCHER service allowing us to understand I/O use patterns by analysing the I/O
use of \textit{all} jobs on the service in a six month period broken down by different
research communities.
The statistics generated by LASSi have been further analysed to gain an understanding of how particular application areas use the file system.  

Our analysis of LASSi I/O data linked to other service data using SAFE allowed us to 
investigate the overall I/O use pattern on ARCHER and has revealed four distinct I/O
use patterns associated with four of the largest research communities on ARCHER:

\begin{itemize}
    \item \textbf{Overall:} The overall I/O use pattern on ARCHER reveals the overlay
    of a range of different patterns with the major ones described below. Over 50\% of
    the use in the analysis period was for jobs that read less than 4~GiB and wrote
    less than 32~GiB. Overall, twice as much data was written than was read on ARCHER
    in the analysis period.
    \item \textbf{Materials science:} Job I/O use is characterised by small amounts of
    data read and written on a \textit{per job} basis but overall high amounts of data
    read and written due to the very large number of jobs. Approximately three times as
    much data was written as was read by the materials science community.
    \item \textbf{Climate modelling:} Job I/O use is characterised by large amounts of
    data read and written on a \textit{per job} basis with a small range of per-job
    read/write behaviours due to the natural constraint of size of scenarios modelled.
    Approximately five times as much data was written as was read by the climate
    modelling community.
    \item \textbf{Computational fluid dynamics:} Job I/O use is characterised by large
    amounts of data read and written on a \textit{per job} basis with a wide range of
    per-job read/write behaviours due to the wide range of sizes of scenarios modelled.
    Approximately five times as much data was written as was read by the CFD community.
    \item \textbf{Biomolecular modelling:} Job I/O use is characterised by small amounts
    of data read and medium amounts of data written on a \textit{per job} basis with a
    wide range of per-job write behaviours due to the variety of modelling scenarios.
    Approximately ten times as much data was written as was read by the biomolecular
    modelling community.
\end{itemize}

Based on our analysis, we were also able to qualitatively predict how the I/O requirements of
each of the communities will change in the future: communities that use grid-based models
(climate modelling, CFD) will see an increase in per-job I/O requirements as the
resolution of the modelling grids increases; the materials science and biomolecular 
modelling would expect to see less change in the per-job I/O requirements (due to
scientific limits on the size of systems to be studied) but would see an overall
increase in I/O requirements as more sophisticated statistical methods and larger
parameter sweeps require more individual jobs per research programme. Future national
services serving these communities will need to take these requirements into account 
in their design and operation.

\section{Future Directions}

We are in the early stages of analysing the data obtained so far and plan to continue our analysis to learn
more about application requirements for I/O.  We expect to find more situations of
applications that do not use the file system in an optimal way. As we find more incidents
of application slowdown we will refine and augment the metrics used by LASSi. We also plan
to automate detection of application slowdown so that we do not have to wait for
individual incident reports to allow us to correlate LASSi metrics and actual incidents
on the system.

We found that the current I/O operations metrics imported into SAFE (total number of I/O ops
over the whole job) are not particularly useful for understanding this aspect of the I/O
use on the system. Importing the peak I/O ops rate (for different operations) for each job
should prove more useful and we plan to develop this functionality so we can analyse
the I/O operations across the service using the powerful combination of LASSi and SAFE
in the same way as we have been able to for data volumes.

This initial analysis has looked at I/O patterns for four of the largest research 
communities on the UK National Supercomputing Service, ARCHER (in terms of core-h use in the
analysis period) but this approach neglects research communities that may have low resource
use overall (measured in core-h) but high or different demands of the I/O resources. We 
plan to modify our analysis to reveal which communities are making different demands of 
the I/O resources by altering the weighting factor for the heatmaps produced from core-h
to both data volume read/written and I/O operations.


We are also working to identify other HPC facilities that routinely collect per-job I/O
statistics to allow us to compare the use patterns on ARCHER and understand how similar
(or different) patterns are for similar communities on different facilities.


In addition to future research directions, we have the following activities planned
to increase the impact and utility of the I/O data and metrics we are collecting:

\begin{itemize}
    \item Integrate LASSi into the data collection framework provided by Cray View for ClusterStor\footnote{\url{https://www.cray.com/products/storage/clusterstor/view}} so that sites with this software can take advantage of the alternative view that LASSi can provide.
    \item Develop an I/O score chart that can be used as part of the ARCHER resource
    request process to give the service a better way to anticipate future I/O requirements
    and improve operational efficiency.
    \item Develop a machine learning model for application run time and its I/O to
    potentially allow the scheduler to make intelligent decisions on how to schedule
    different job types to reduce I/O impact between jobs and on the wider service.
\end{itemize}
\section*{Acknowledgments}

This work used the ARCHER UK National Supercomputing Service. We would like to acknowledge EPSRC, EPCC, Cray, the ARCHER helpdesk and user community for their support.

\bibliographystyle{IEEEtran}

\bibliography{IEEEabrv,paper.bib}

\end{document}